\documentclass[journal]{IEEEtran}

\usepackage[utf8]{inputenc}
\usepackage{amsmath, amssymb, amsthm}
\usepackage{enumerate}
\usepackage{graphicx}
\usepackage{xcolor}
\usepackage{booktabs}
\usepackage[utf8]{inputenc} 
\usepackage[T1]{fontenc}
\usepackage{url}   
\usepackage[noadjust]{cite}

\usepackage{stfloats}
\usepackage {tikz}
\usepackage{pgfplots}
\pgfplotsset{compat=newest,compat/show suggested version=false}
\usetikzlibrary {shapes,arrows,positioning}
\usetikzlibrary {arrows.meta} 
\usepackage{multirow}
\usepackage{color, colortbl}
\definecolor{Gray}{gray}{0.9}
\usepgfplotslibrary{fillbetween} 
\usepackage{bm}
\makeatletter
\newcommand{\ALG@lineautorefname}{Algorithm}
\makeatother

\usepackage{subfig}

\usepackage{setspace}

\usepackage{algorithm, algorithmicx, algpseudocode}
\algrenewcommand\algorithmicrequire{\textbf{Input:}}
\algrenewcommand\algorithmicensure{\textbf{Output:}}
\algnewcommand{\BlackBox}[1]{%
    \begin{flushleft}
    \hspace{-.7cm}
    \textbf{Available Functions:}
    {\raggedright #1}
    \end{flushleft}
}
\algnewcommand{\Initialize}[1]{%
    \begin{flushleft}
    \hspace{-.7cm}
    \textbf{Initialize:}
    {\raggedright #1}
    \end{flushleft}
}

\usepackage{hyperref}
\usepackage[capitalize,nameinlink]{cleveref}
\usepackage{centernot}
\usepackage[normalem]{ulem}

\usepackage{color, soul}

\usepackage{mdframed}
\mdfdefinestyle{MyFrame}{%
    linecolor=black,
    outerlinewidth=2pt,
    roundcorner=20pt,
    innertopmargin=\baselineskip,
    innerbottommargin=\baselineskip,
    innerrightmargin=10pt,
    innerleftmargin=10pt,
    backgroundcolor=gray!20!white
}

\usepackage{acronym}
\newacro{mimo}[MIMO]{multiple-input multiple-output}

\newcommand{\CC}{{\mathcal C}}
\newcommand{\CD}{{\mathcal D}}
\newcommand{\CF}{{\mathcal F}}

\newcommand{\CP}{{\mathcal P}}
\newcommand{\CS}{{\mathcal S}}

\newcommand{\fpkq}{\CF_{p}(k, q)}
\newcommand{\fpkp}{\CF_{p}(k, p)}

\newcommand{\BC}{{\mathbb C}}
\newcommand{\BE}{{\mathbb E}}
\newcommand{\BF}{{\mathbb F}}

\newcommand{\fp}{\BF_{p}}
\newcommand{\fq}{\BF_{q}}

\newcommand{\fqx}{\BF_{q}[X]}

\newcommand{\dH}{d_{{\rm H}}}

\newcommand{\dc}{d_{{\rm C}}}
\newcommand{\rhoh}{\rho_{{\rm H}}}
\newcommand{\rhos}{\rho_{{\rm C}}}
\newcommand{\rhosbar}{Q_{\rm e}}

\newcommand{\herm}[1]
{{#1}^{\dagger}}

\newcommand{\abs}[1]
{{\raisebox{-0.25\depth}{$\biggl\lvert$}}{#1}\raisebox{-0.25\depth}{$\biggr\rvert$}}
\newcommand{\norm}[1]
{\left\|{#1}\right\|}

\DeclareMathOperator{\CCP}{CP}

\DeclareMathOperator{\GRS}{GRS}

\DeclareMathOperator*{\argmax}{arg\,max}

\DeclareMathOperator*{\expec}{\mathbb{E}}
\DeclareMathOperator*{\var}{var}

\newtheorem{theorem}{Theorem}
\newtheorem{lemma}{Lemma}

\newtheorem{corollary}{Corollary}

\theoremstyle{definition}
\newtheorem{definition}{Definition}

\theoremstyle{plain}
\newtheorem{remark}{Remark}

\newcommand{\hide}[1]
{{\iffalse #1 \fi}}

\definecolor{darkbrown}{RGB}{101,67,33}
\usepackage{color}

\begin{document}

\title{Precoding Design for Limited-Feedback {MIMO} Systems via Character-Polynomial Codes} 

\IEEEoverridecommandlockouts
\author{%
    \IEEEauthorblockN{Siva Aditya Gooty,~\textit{Student Member, IEEE},
Samin Riasat,~\textit{Student Member, IEEE},
\\Hessam Mahdavifar,~\textit{Member, IEEE},
and Robert W. Heath, Jr.,~\textit{Fellow, IEEE}}
        \thanks{S. A. Gooty, S. Riasat, and H. Mahdavifar are with the 
        Department of Electrical and Computer Engineering, Northeastern 
        University, Boston, MA 02115, USA (e-mail: gooty.s@northeastern.edu; 
        riasat.s@northeastern.edu; h.mahdavifar@northeastern.edu).}
        \thanks{R. W. Heath, Jr. is with the Department of Electrical and 
        Computer Engineering, University of California San Diego, 
        La Jolla, CA 92122, USA (e-mail: rwheathjr@ucsd.edu).}
        \thanks{This paper was presented in part at the 2025 IEEE International Conference on Communications (ICC) \cite{gooty2025precodingdesignlimitedfeedbackmiso}.}
        \thanks{This work was supported in part by NSF under Grants CCF-2415440, CCF-2435254 and CNS-2433782 and in part by the Center for Ubiquitous Connectivity (CUbiC) under the JUMP 2.0 program.}
}
\markboth{SUBMITTED TO IEEE TRANSACTIONS ON COMMUNICATIONS, JULY 2026}%
{SUBMITTED TO IEEE TRANSACTIONS ON COMMUNICATIONS, JULY 2026}
\maketitle

\vspace{-20mm}

     \begin{abstract}
This paper presents a precoding codebook design for limited-feedback multiple-input
multiple-output (MIMO) systems under the equal-gain transmission (EGT) constraint. In particular, we demonstrate that character--polynomial (CP) codes provide a structured solution that achieves constant-envelope transmission, low storage complexity, and Grassmannian packing without dependence on array geometry or channel statistics. In contrast to geometry-dependent discrete Fourier transform (DFT) codebooks used in current 5G systems and unstructured Grassmannian codebooks with high storage complexity, CP codebooks combine practical implementation advantages with near-optimal packing performance. For multiple-input single-output (MISO) systems, we derive an upper bound on the mean squared quantization error and show that the distortion relative to the EGT baseline vanishes asymptotically as the number of transmit antennas increases and the code rate approaches one. For MIMO systems with two receive antennas, we develop an iterative method to establish the EGT baseline. Simulation results under Rayleigh, correlated, and clustered delay line (CDL) channel models show that CP codebooks approach the EGT baseline across all channel conditions while outperforming phase shift keying (PSK) and 5G DFT codebooks in several operating regimes, and, in the Rayleigh fading case, incur negligible packing loss relative to numerically optimized Grassmannian codebooks obtained via the alternating projection (AP) method.


\end{abstract}

\begin{IEEEkeywords}
MIMO, Precoding designs, Grassmann packing. 
\end{IEEEkeywords}

\section{Introduction}
\label{sec:introduction}

\subsection{Background and motivation}

MIMO communication improves spectral efficiency and link reliability through spatial multiplexing and beamforming~\cite{bjornson20246gmimomassivespatial, heath2018foundations}. These gains can be further improved when channel state information (CSI) is available at the transmitter, enabling precoding and rate adaptation
{\cite{heath2018foundations}.} 
The most common means of obtaining CSI at the transmitter is limited feedback, whereby the receiver quantizes the CSI and conveys it to the transmitter over a finite-rate feedback channel~\cite{Heath_1998_partialfeedback, Love_2008_overview, albreem2021overview}. More specifically, limited-feedback precoding requires the transmitter to select 
a beamforming vector from a finite codebook, and the codebook 
design determines how well the quantized vector approximates 
the optimal beamforming direction, see, e.g.,~\cite{Love_2008_overview} for a comprehensive overview of limited-feedback techniques and ~\cite{albreem2021overview}
for a survey of precoding methods in MIMO systems.
The problem of limited feedback for precoding in MIMO systems is sometimes known as Grassmannian feedback, due to the mathematical connections between quantizing the dominant subspaces of a MIMO channel and packing problems in the Grassmannian space \cite{love2003grassmannian}. In particular, for rank-1 precoding, codebook design reduces to the 
problem of packing lines in the Grassmannian space.

While Grassmannian codebooks provide strong beamforming performance, they are often unstructured and require large storage complexity. As a result, practical systems such as 5G New Radio (NR) rely on structured beam codebooks based on DFT beams. 
{In particular, 
the standardized 
Type~I and
Type~II codebooks employ 
an oversampled DFT beam 
grid together with additional system-level enhancements, providing a practical constant-envelope implementation with low complexity~\cite{Ning_2006_dft}.} Such beam-based codebooks perform well when the channel characteristics align with the underlying beam structure, but may experience performance degradation in rich scattering or geometry-mismatched environments. More generally, practical precoding codebooks must simultaneously satisfy multiple requirements, including constant-envelope transmission, low storage complexity, and robustness across diverse channel conditions. This suggests a gap between unstructured Grassmannian codebooks with near-optimal packing and practical structured beam codebooks. Consequently, designing structured codebooks that achieve good Grassmannian packing while satisfying practical implementation constraints remains an open design challenge.

\subsection{Related work}

The application of Grassmannian packing to precoding codebook design for limited-feedback MIMO systems has been extensively studied in the literature. Early work established bounds on beamforming gain as a function of the minimum chordal distance between codewords and showed that codebooks designed via Grassmannian line packing achieve near-optimal performance~\cite{love2003grassmannian}. Resulting codebooks were obtained either from known algebraic constructions~\cite{love2003grassmannian} or through numerical optimization methods such as AP~\cite{dhillon2008constructing}. Recognizing that codebooks designed for independent and identically distributed (i.i.d.) channels may be suboptimal in correlated channel conditions, subsequent works extended Grassmannian codebook design to spatially correlated channels by skewing or rotating the codebook on the Grassmann manifold based on channel covariance information~\cite{love2004grassmannian}. These ideas were further extended to multi-antenna broadcast channels~\cite{Choi2013Raghavan} and time-varying channels~\cite{schwarz2013adaptive}. However, such approaches require knowledge of channel statistics at the transmitter and must be redesigned or adapted when the channel distribution changes.

From a practical implementation perspective, EGT was proposed as 
a phase-only beamforming technique that achieves full diversity using only quantized phase information at the transmitter, establishing a practical performance benchmark for constant-envelope codebook designs~\cite{Love_2003_EGT,Murthy_EGT,Tsai_EGT}. Constellation-based codebooks 
were subsequently proposed for rank-1 beamforming, eliminating explicit codebook storage by exploiting constellation structure, though at the cost of suboptimal Grassmannian packing~\cite{Ryan_2009_psk}. In practical cellular systems such as 5G, codebooks based on the 
DFT are widely used due to their constant-modulus structure and hardware-friendly implementation. DFT codebooks perform well in spatially correlated channels with uniform linear array (ULA) geometry~\cite{Yangdft2010, Ning_2006_dft}, and have been extended to 
multi-panel arrays~\cite{Fu_2022_dft}. However, DFT codebooks are inherently tied to specific antenna geometries and spatial correlation structures, and their performance may degrade in rich scattering environments under more general array configurations. 

{Structured alternatives such as Kerdock codes~\cite{Inoue_2009_kerdock}
and trellis-based designs~\cite{au2011trellis, choi2013noncoherent}
exploit algebraic or coding-theoretic structure to reduce storage and
search complexity while providing practical constant-envelope codebooks.}
$\CCP$ codes were subsequently introduced
in
{\cite{Soleymani_2020_subspace, Soleymani_2022_subspace, soleymani2021new}
}as a class of
structured analog subspace codes for non-coherent wireless communication,
providing a new algebraic construction of Grassmannian subspace codes.
Decoding algorithms for these codes were later developed
in
{\cite{riasat2024decodinganalogsubspacecodes}}. In this work, we
leverage $\CCP$ codes to design precoding codebooks for limited-feedback
MIMO systems. The resulting codebooks retain the constant-envelope
property, require low storage complexity due to their algebraic
structure, and can be used without modification across different channel
models without requiring channel statistics or codebook rotation.

\subsection{Contributions}

Building on the preliminary MISO results in~\cite{gooty2025precodingdesignlimitedfeedbackmiso}, 
{this paper focuses on rank-1 precoding and considers 
MISO systems and MIMO systems with two receive 
antennas.} 
The main contributions of this work are summarized as follows:

\begin{itemize}
    \item \textbf{Performance analysis of CP codebooks:}
    We analyze the beamforming performance of 
    $\CCP$ codebooks for limited-feedback precoding and derive an upper bound on the mean squared quantization error for MISO systems. The analysis shows that the normalized distortion with respect to the 
    EGT baseline vanishes asymptotically as the number of transmit antennas increases and the code rate approaches unity.

    \item \textbf{EGT baseline for MIMO systems:}
    For MIMO systems with multiple receive antennas, we develop an iterative algorithm to compute the EGT baseline for equal-gain codebooks. This provides a practical performance benchmark for constant-envelope precoding in MIMO systems where a closed-form characterization of the EGT gain is not available.

\item \textbf{Channel-agnostic structured codebook design:}
We show that $\CCP$ codebooks can be used for precoding without requiring knowledge of channel statistics or array geometry at the transmitter. 
{We evaluate the performance of a fixed $\CCP$ codebook across i.i.d. Rayleigh, spatially correlated, and CDL channel models without any channel-dependent adaptation.} The same structured codebook is used in all cases, without requiring channel-dependent codebook rotation, while retaining the constant-envelope property and low storage complexity.

\item \textbf{Comparison with practical codebooks:}
{We provide a comprehensive performance comparison with DFT and PSK codebooks, which represent the most relevant constant-envelope baseline.} The results show that DFT performs well in highly correlated channels at low feedback rates, while $\CCP$ codebooks outperform DFT at moderate and high feedback rates and consistently outperform PSK codebooks, achieving up to approximately 1~dB gain. These results highlight the tradeoff between geometry-dependent and channel-agnostic codebook designs and demonstrate the flexibility of $\CCP$ codebooks in trading feedback rate for beamforming gain.
\end{itemize}

The rest of the paper is organized as follows.
\autoref{sec:Preliminaries} reviews the necessary background
on Grassmannian geometry, subspace codes, and the DFT and
PSK codebook constructions.
\autoref{sec:System model} presents the system model for
limited-feedback beamforming under the equal-gain constraint
for both MISO and MIMO systems.
\autoref{sec:iter method} describes the CP
codebook construction, analyzes its storage complexity, and derives upper bounds on the mean
squared quantization error. It also introduces an iterative algorithm to compute the EGT baseline
for MIMO systems and extends the design to correlated channel conditions. 
\autoref{sec:Simulation} presents simulation results under i.i.d. Rayleigh, spatially correlated Rayleigh, and standardized CDL channel models for both MISO and MIMO systems, including comparisons with DFT and PSK codebooks and remarks on the role of channel structure in codebook performance.~\autoref{sec:Future Work} summarizes the main findings and discusses directions for future work.

\section{Preliminaries}
\label{sec:Preliminaries}

In limited-feedback precoding, beamforming vectors are selected from a finite codebook, and the performance depends on how well the selected vector aligns with the channel direction. This problem can be naturally formulated in terms of subspace quantization on the Grassmann manifold, where distances between subspaces determine beamforming performance. In this section, we briefly review the necessary concepts from Grassmannian geometry and analog subspace coding that will be used in the sequel.

 \subsection{Notation conventions}

We use bold capital letters (e.g. $\mathbf{X}$) to represent matrices, and bold small letters (e.g. $\mathbf{x}$) for 
vectors. We also reserve $|\cdot|$ for the absolute value of a complex number and use $\norm{\,\cdot\,}$ for the $L^{2}$-norm of a vector. 
{For a complex vector, $(\cdot)^{*}$ denotes complex
conjugation, $(\cdot)^{\text{T}}$ denotes transpose, and $(\cdot)^{\dagger}$
denotes conjugate transpose (Hermitian transpose). Complex scalars use the same conjugation notation $(\cdot)^{*}$.}

\subsection{Chordal distance and subspace codes}

Given an ambient vector space $W$\hide{ with elements regarded as row vectors}, 
the set $\CP(W)$ denotes all subspaces of $W$, and $\CP_{m}(W)$ denotes the set of all $m$-dimensional subspaces. 
$\CP_{m}(\BC^{n})$ is referred to as a \emph{Grassmannian space} and is denoted $G_{m, n}(\BC)$. \hide{The elements of $G_{m, n}(\BC)$ are called \emph{$m$-planes}. }Any \hide{$m$-plane }$U \in G_{m, n}(\BC)$ is equipped with the natural inner product $\langle {\bf u}, {\bf v}\rangle := \herm{{\bf u}} {\bf v}$ for ${\bf u}, {\bf v} \in U$. 

\begin{definition}
    \label{def:chordal}
    Let $U, V \in G_{m, n}(\BC)$ \hide{be $m$-planes}. 
    Let $\mathbf{u}_i \in U$ and $\mathbf{v}_i \in V$, for $i \in [m]$, be unit vectors that form orthonormal bases, with each $\mathbf{u}_i$ a basis vector for $U$ and each $\mathbf{v}_i$ a basis vector for $V$ chosen recursively to maximize $|\langle \mathbf{u}_i, \mathbf{v}_i \rangle|$.
    The $i$-th principal angle $\theta_i$ between $U$ and $V$ is defined as $\theta_i = \arccos |\langle \mathbf{u}_i, \mathbf{v}_i \rangle|$. Then the \emph{chordal distance} \cite{Conway96} between $U$ and $V$ is 
    \begin{align}
        \label{eq:dc}
        \dc(U, V) 
        := \sqrt{\sum_{i = 1}^{m} \sin^{2} \theta_{i}}
    \end{align}
    An \emph{analog subspace code} is a collection  $\CC \subseteq \CP(\BC^{n})$ of subspaces. 
\end{definition}

\subsection{Grassmann codes}
In the context of limited-feedback precoding, a Grassmann code corresponds to a precoding codebook whose codewords represent beamforming subspaces, and the design objective is to maximize the minimum chordal distance between codewords. In fact, a \emph{Grassmann code} can be considered as an analog subspace code where the dimension of all of its codewords are equal. 
Shannon~\cite{shannon1959probability} 
and Barg and Nogin~\cite{barg2002bounds, barg2006bound} 
established bounds on the minimum distance of Grassmannian 
codes. Conway et al.~\cite{Conway96} and Henkel~\cite{henkel2005sphere} 
characterized packings in Grassmannian spaces, while Dhillon 
et al.~\cite{dhillon2008constructing} developed constructive 
packing methods via AP. Shor and 
Sloane~\cite{shor1998family} and Calderbank et 
al.~\cite{calderbank1999group} proposed structured codebook 
constructions with algebraic properties. The $\CCP$ codes used 
in this paper belong to the class of structured analog 
subspace codes~\cite{Soleymani_2020_subspace, Soleymani_2022_subspace} 
and are described next.
\subsection{CP codes}

Let $\fq$ be a finite field of size $q$ and characteristic $p$. For $k < q$, Let 
    \begin{align}
        \label{eq:fkq}
        \CF(k, q) 
        &:= \{f \in \fqx: \deg(f) \le k\} 
    \end{align}
denote the set of all polynomials of degree at most $k$ over $\fq$. This set serves as the message space for classical Reed--Solomon (RS) codes. The elements of $\CF(k, q)$ are called \emph{message polynomials}, whose coefficients represent the message symbols. 
\hide{For 
$f \in \CF(k, q)$, we let 
    $f^{(p)}(X) 
    := \sum_{0 \le j \le k / p} 
    f_{j p} X^{j}$
and}We also define \hide{the \hide{additional }message spaces $\fpkq$ and $\fpkq'$ as follows. }\hide{
\begin{align} 
    \fpkq 
    &:= \{f(X) - f^{(p)}(X^{p}): f \in \CF(k, q)\} 
    \label{eq:fpkq} 
    \\ 
    \fpkq' 
    &:= \{f(X) / X: f \in \fpkq\}.
    \label{eq:fpkq'}
\end{align}
In other words, 
}$\fpkq$ to be the set of all $f(X) = \sum_{j} f_{j} X^{j} \in \CF(k, q)$ with $f_{j p} = 0$ for all integers $j \ge 0$, 
and $\fpkq' := \{f(X) / X: f \in \fpkq\}$. 

\begin{definition}[$\CCP$ Code {\cite[Definition~6]{Soleymani_2022_subspace}}]
    \label{def:cp}
    {A character of $\fq$ is a group homomorphism $\chi : (\fq, +) \to \mathbb{C}$ satisfying $\chi(a + b) = \chi(a)\chi(b)$ for all $a, b \in \fq$, mapping field elements to complex numbers on the unit circle, and  non-trivial if $\chi \not\equiv 1$.}
    Fix $k \le n < q$, a non-trivial character $\chi$ of $\fq$, and units $\alpha_{1}, \dots, \alpha_{n} \in \fq^{\times} := \fq \setminus \{0\}$. Then the encoding of $f \in \fpkq$ in $\CCP := \CCP_{n}(\fpkq, \chi) \subseteq G_{1, n}(\BC)$ is given by
    \begin{align}
        \label{eq:cp}
        \CCP(f) 
        &:= (\chi(f(\alpha_{1})), \dots, \chi(f(\alpha_{n}))), 
    \end{align}
    where we identify $\CCP(f)$ with the one-dimensional subspace $\langle \CCP(f) \rangle$. 
\end{definition}
Note that the resulting $\CCP$ codewords are constant-modulus complex vectors whose entries lie on the unit circle, making them suitable for equal-gain transmission and hardware-efficient beamforming implementations.


Given $k \le n \le q$, distinct $\alpha_{1}, \dots, \alpha_{n} \in \fq$ and 
not necessarily distinct $\beta_{1}, \dots, \beta_{n} \in \fq^{\times}$, the encoding of $f \in \CF(k - 1, q)$ in $\GRS := \GRS_{n}(\CF(k - 1, q))$ is given by
\begin{align}
    \label{eq:grs-definition}
    \GRS(f) 
    &:= (\beta_{1} f(\alpha_{1}), \dots, \beta_{n} f(\alpha_{n})).
\end{align}
Let $\beta_{i} = \alpha_{i}$ for $i \in \{1, \dots, n\}$. 
Using the notation from~\autoref{def:cp}, 
prior to concatenation by $\chi$, $\CCP_{n}(\fpkq, \chi)$ can be expressed as $\GRS_{n}(\fpkq')$. 
Therefore, 
\begin{align}
    \label{eq:CPsize}
    |\CCP| = |\fpkq| 
    = |\fpkq'| = q^{k - \lfloor k / p \rfloor}.
\end{align}
See also \cite[Theorem~9]{Soleymani_2022_subspace}.

For the sake of simplicity, our primary focus will be on the case of prime fields, i.e., when $q = p$. 
In this case, observe that the dimension of $\CCP_{n}(\CF_{p}(k, q), \chi)$, which is $k$ by \eqref{eq:CPsize}, is equal to the dimension of $\GRS_{n}(\CF(k - 1, q))$, where the message polynomials have degree at most $k - 1$. 
\subsection{Covering radius and mean quantization error
}
\label{sec:covering}

The \emph{covering radius} of a block code $\CC \subseteq \BF^{n}$ is defined as 
\begin{align}
    \label{eq:rho}
    \rhoh(\CC) 
    &:= \max_{{\bf y} \in \BF^{n}} 
    \min_{{\bf c} \in \CC} 
    \dH({\bf y}, {\bf c}), 
\end{align}
where 
$\dH$ is the Hamming distance. 
For instance, $\rhoh(\GRS) = n - k$ (see, e.g.~\cite{Huffman03}). 
The chordal covering radius of an analog subspace code $\CC \subseteq \CP(\BC^{n})$ may be analogously defined as 
\begin{align}
    \label{eq:rhos}
    \rhos(\CC) 
    &:= \max_{U \in \CP(\BC^{n})} 
    \min_{V \in \CC} 
    \dc(U, V),
\end{align}
where $\dc(.,.)$ is the chordal distance defined in \eqref{eq:dc}.

%
The covering radius can be thought of as the \textit{maximum error}, in the sense of the underlying distance defined for the space, when using a code for the quantization of the space. 
Note that covering radius characterizes the worst-case scenario of the quantization process. In practice, however, given a certain distribution over the space, the average quantization error becomes more relevant. This is defined formally next. 
\begin{definition}[Mean Squared Quantization Error]
\label{meancoverdef}
    Let $\CC \subseteq \CP(\BC^{n})$ be an analog subspace code and $\CD$ a distribution on $\CP(\BC^{n})$. Then the \emph{mean squared quantization error} of $\CC$ over $\CD$ is defined as 
    \begin{align}
        \label{eq:rhosbar}
        \rhosbar(\CC) 
        &:= \expec_{U \sim \CD} 
        \left[\min_{V \in \CC} 
        \dc(U, V)^2\right].
    \end{align}
\end{definition}
Note also that $\rhosbar(\CC) \le \rhos(\CC)^2$.
\hide{
\subsection{Related Work}
\label{sec:Recent works}

The precoding design at the transmitter continues to be a primary research focus in 6G MIMO network system design. 
Schwarz et al.~\cite{Schwarz_2019_grassmannian} construct dual stage channel quantization schemes for FDD channels to address the accurate high dimensional CSI estimation issue via product codebook. 
Jang et al.~\cite{Jang_2023_Prob_MIMO} address the channel dimension issues with techniques to quantize the CSI at the receiver and reconstruct the different groups of channel elements with the help of pre-defined probability distribution. Ni et al.~\cite{Ni_2023_subspace_distortion} maximize the sum rate for mmWave communication using hybrid precoding design considering both baseband and passband quantization. 
Kim et al.~\cite{Kim_2022_IRSmimo} put forward solutions to maximizing the data rate in intelligent reflecting surface (IRS) aided communications between BS and UE. The codebook mechanisms for capacitance vectors via deep reinforcement learning techniques consider the channel estimation problem and limited data rate of the feedback. Zhou et al.~\cite{Zhou_2024_mmse} propose robust channel quantization techniques in both iterative and non iterative fashions to address the quantization errors when the feedback bits is not sufficient using weighted minimum mean square and minimum mean square error precoding algorithms respectively. Jang et al.~\cite{Jang_2022_DLmimo} design a deep learning framework to unify the channel acquisition at UE and precoding design at BS for multi user MIMO systems. 
Turan et al.~\cite{Turan_2024_fdd} propose Gaussian mixture model (GMM) with reduced parameters to account for all UEs in a specific BS cell. The analytical strategy using GMM lowers the complexity of determining the feedback at UE and supports parallelization in multi user MIMO systems. Extremely large MIMO (XL-MIMO) systems have recently been extensively investigated as a potential success for 6G networks \cite{Wang_2024_xlmimo}. 
For the XL-MIMO systems, Dreifuerst and Heath~\cite{Dreifuerst_2024_HeirMIMO} present a new algorithm with neural network and a beam selection strategy that demonstrate improved performance in multi-path time domain channels when compared to conventional 5G DFT codebooks.
}


Before presenting the system model and $\CCP$ codebook construction, we briefly describe the DFT and PSK codebook constructions. Both satisfy the equal-gain condition and are compared against the proposed $\CCP$ codebook in~\autoref{sec:Simulation}.
\subsection{DFT codebook}

DFT-based codebooks are widely used in practical systems due to their constant-envelope structure and compatibility with ULA antenna architectures~\cite{Ning_2006_dft,Yangdft2010}. In 5G 
NR, Type~I and Type~II codebooks are based on DFT beamforming and incorporate additional features such as dual polarization and beam-group selection~\cite{Ning_2006_dft}. 
{In particular, 
both Type~I and Type~II codebooks select beams from
the same oversampled beam grid,
which can be viewed as an oversampled DFT (or equivalently, a sampled array-response/DTFT) dictionary. Type~I selects a single beam, while Type~II selects and combines multiple beams. This oversampled grid coincides with an orthogonal
DFT matrix only when the oversampling factor $O_1 = 1$. For $O_1 > 1$, it is
a sampled DTFT of the array
response. Since this work focuses on rank-1 precoding under the equal-gain constraint, we use the corresponding oversampled DFT beam codebook
(i.e., single-beam
selection as in Type~I)
as the practical baseline,
independent of 
dual polarization, beam-group
selection, and Type~II's multi-beam combining.}

For $N_{\text{T}}$ transmit antennas, the oversampled DFT codebook with oversampling factor 
$O_1$ consists of $N_{\text{T}}O_1$ codewords. The $\ell$-th codeword is given by
\begin{equation}
\mathbf{d}_{\ell}
=
\frac{1}{\sqrt{N_{\text{T}}}}
\left[
1,\;
e^{j2\pi \ell/(N_{\text{T}}O_1)},\;
\ldots,\;
e^{j2\pi (N_{\text{T}}-1)\ell/(N_{\text{T}}O_1)}
\right]^{\text{T}},
\end{equation}
for $\ell\in\{0,1,\ldots,N_{\text{T}}O_1-1\}$. Each codeword satisfies the equal-gain constraint since all entries have magnitude $1/\sqrt{N_{\text{T}}}$. The codebook size is
$|\mathcal{C}_{\text{DFT}}|=N_{\text{T}}O_1$, requiring
$\lceil\log_2(N_{\text{T}}O_1)\rceil$ feedback bits. Increasing the oversampling factor $O_1$ refines the beamforming resolution and increases the achievable feedback rate.

\subsection{PSK codebook}

Constellation-based codebooks using PSK alphabets were proposed in~\cite{Ryan_2009_psk} for limited-feedback beamforming. In the PSK codebook, each entry of the beamforming vector is 
{$1/\sqrt{N_\text{T}}$ times an element of}
the $M$-PSK alphabet $\{e^{j2\pi m/M}: m=0,\ldots,M-1\}$, and the first entry is fixed to $1/\sqrt{N_{\text{T}}}$ without loss of generality. The remaining $N_{\text{T}}-1$ entries each independently take $M$ values, yielding a codebook of size $|\mathcal{C}_{\text{PSK}}|=M^{N_{\text{T}}-1}$. The number of feedback bits is $\lceil (N_{\text{T}}-1)\log_2 M\rceil$, which grows linearly with both the number of transmit antennas and the constellation order.



\section{System Model}
\label{sec:System model}

This section defines the system model for limited-feedback beamforming in a MIMO system. The CSI at the receiver (CSIR) is assumed to be perfect and the transmitter selects a beamforming
vector from a finite codebook based on quantized feedback. The objective is to design a beamforming codebook that maximizes the effective channel gain under a finite feedback constraint and a constant-envelope (equal-gain) transmission constraint. In the MIMO system considered, shown in~\autoref{fig:System model}, the transmitter 
is equipped with $N_{\text{T}}$ transmit antennas and the receiver with $N_{\text{R}} = 2$ antennas. The receiver with perfect CSIR enables the transmitter to compute the optimal beamforming vector. 
Let the channel $\mathbf{H} \in \mathbb{C}^{N_{\text{R}} \times N_{\text{T}}}$ be a memoryless complex fading channel with i.i.d. entries. With transmit beamforming and receive combining, let $x\in \mathbb{C}$ and $y\in \mathbb{C}$ denote the input and output of the system with 
\begin{align}
\label{eqn:System model}
    y = \mathbf{z}^\dagger \mathbf{H}\mathbf{f}x + \mathbf{{z^\dagger}{n}},
\end{align}
where $\mathbf{f} \in \mathbb{C}^{N_{\text{T}} \times 1} $ is the beamforming vector with $\|\mathbf{f}\| = 1$, $\mathbf{z} \in \mathbb{C}^{N_{\text{R}} \times 1}$ is the combining vector and $\mathbf{n} \in \mathbb{C}^{N_{\text{R}} \times 1}$ is a circularly symmetric complex Gaussian random vector with variance $N_0$. 
Then the overall gain $\Gamma(\mathbf{H})$, given the beamforming and combining vectors of the system, 
is given by
\begin{align}
\label{eqn:beamforming gain}
    \Gamma(\mathbf{H}) = |\mathbf{z}^{\dagger}\mathbf{H}\mathbf{f}|^2. 
\end{align}
Ideally, maximum ratio transmission (MRT) and maximum ratio combining (MRC) are employed to maximize the overall gain~\cite{Tse_Viswanath_2005}.
This is done by setting the combining vector to the left singular vector and beamforming vector to the right singular vector of the channel. The MRT beamforming vector maximizes the received signal power without any constraint on the amplitude of the beamforming weights. 

Many practical systems require constant-envelope or equal-gain transmission to enable
efficient power amplifier operation. Under this constraint, the optimal beamforming strategy is EGT, 
which serves as a practical performance benchmark for constant-envelope codebook design. Then the codebook design problem reduces to designing a set of equal-gain beamforming vectors that quantize the optimal beamforming direction with respect to chordal distance. Therefore, in this work, we evaluate codebook performance relative to the EGT gain.

This paper addresses two system configurations: a MISO system with a single receiver
antenna ($N_{\text{R}} = 1$) and a MIMO system with two receiver antennas ($N_{\text{R}} = 2$).
\subsection{MISO system}
\label{sec:MISO system}
\begin{figure}[!t]
    \centering
    \includegraphics[width=1.0\linewidth, height = 7.0cm]{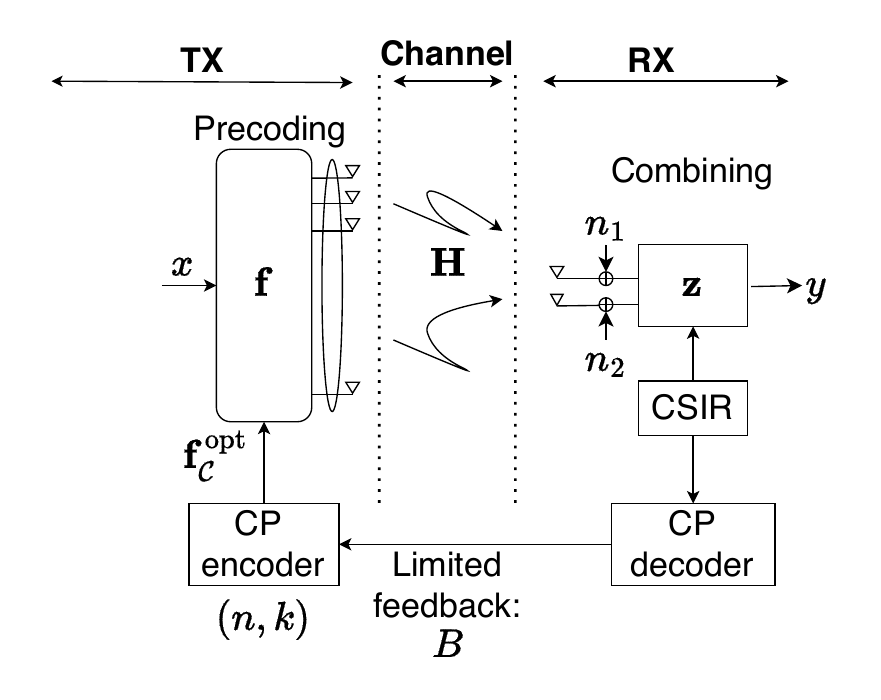} 
    \caption{Block diagram of the limited-feedback MIMO beamforming system. The transmitter does precoding with a beamforming vector $\mathbf{f}$ selected from a $\CCP$ codebook. The receiver computes the
optimal codeword from its CSIR and feeds the index back using $B 
$ bits. The $\CCP$ encoder and decoder share only the evaluation points and character function,
avoiding explicit codebook storage.}
    \label{fig:System model}
\end{figure}

For $N_{\text{R}}=1$, the transmitter receiver model in~\autoref{fig:System model} can be considered as a special case MISO system with channel $\mathbf{h}$ under consideration~\cite{gooty2025precodingdesignlimitedfeedbackmiso}. Note that the system model representation in~\eqref{eqn:System model} now becomes
\begin{align}
\label{eqn:System model MISO}
    y = {z}^* 
{\mathbf{h}^{\text{T}}}\mathbf{f}x + z^*{n}.
\end{align}
{Here $\mathbf{h} \in \mathbb{C}^{N_\text{T} \times 1}$ denotes the channel
coefficients as a column vector,}
with $z$ being the combining unit complex scalar and $n$ being a circularly symmetric complex Gaussian random variable with variance $N_0$. 
{The channel coefficients $\mathbf{h}$ are drawn from a
distribution $\CD$ on $\mathbb{C}^{N_{\text{T}}}$.}

The goal is to design a codebook $\mathcal{C}$ for the beamforming vectors that the transmitter and receiver can agree upon prior to communication. Note that $\mathbf{h}$ is assumed to be only known
at the receiver. Hence, only the receiver (and not the transmitter) can compute the combining and beamforming vectors. 
 The receiver then maps the optimal beamforming vector {$\mathbf{f^{\rm opt}} = \gamma 
{\mathbf{h^*}}/\|\mathbf{h}\|$,} where $\gamma$ is a unit complex number, 
to the \textit{closest} (to be specified later) element of the codebook. The receiver feeds back only the index of the selected codeword using $B = \lceil \log_2 |\mathcal{C}| \rceil$ bits, and the transmitter uses the corresponding beamforming vector for transmission. 


The MRT gain, denoted 
by $\Gamma_\textup{MRT}$, is equal to $\Gamma_\textup{MRT}(\mathbf{h}) = \|\mathbf{h}\|^2$, when 
the combining unit complex scalar, {$ z = 
{\mathbf{h}^{\text{T}}} f/\|
{\mathbf{h}^{\text{T}}} f\|$.}
Love et al.~\cite{love2003grassmannian} and Ryan et al.~\cite{Ryan_2009_psk} showed that the beamforming gain of the codebook $\mathcal{C}$ relates to $\Gamma_\textup{MRT}(\mathbf{h})$ as follows:
\begin{equation}
\label{codegain}
\frac{\Gamma_{\mathcal{C}}(\mathbf{h})}{\Gamma_\textup{MRT}(\mathbf{h})} = \cos^2\theta(\mathbf{f^{\rm opt}},\mathbf{f^{\rm opt}_{\mathcal{C}}}),
\end{equation}
where $\mathbf{f^{\rm opt}_{\mathcal{C}}} \in \mathcal{C}$ maximizes the beamforming gain, and $\theta(.,.)$ denotes the principal angle. 
In other words, the problem of searching for the beamforming vector in $\mathcal{C}$ is equivalent to solving the following:
\begin{align}
\label{eqn:closet angle}
    \mathbf{f^{\rm opt}_{\mathcal{C}}} = \argmax_{\mathbf{f} \in \mathcal{C}} \cos^2\theta(\mathbf{f^{\rm opt}},\mathbf{f}). 
\end{align}

\begin{remark}
    \label{rmk:maximization}
    The maximization in \eqref{eqn:closet angle} (equivalent to maximizing the beamforming gain in \eqref{codegain}) 
    is equivalent to minimizing the chordal distance between $\mathbf{f}^{\rm opt}$ and $\mathbf {f}$. This is established in \cite{love2003grassmannian}, 
    relating the problem of beamforming design and packing lines in the Grassmannian space. Then, given the codebook $\mathcal{C}$
    , the problem to solve at the receiver is to search for the codeword that is closest, in chordal distance, to $\mathbf{f}^{\rm opt}$. 
\end{remark}

\begin{remark}
    \label{rmk:qerror}
    In light of \autoref{rmk:maximization}, note also that
    \begin{equation}
\label{quantization_gain}
\dc(\mathbf{f^{\rm opt}},\mathbf{f^{\rm opt}_{\mathcal{C}}})^2 = 1 - \cos^2\theta(\mathbf{f^{\rm opt}},\mathbf{f^{\rm opt}_{\mathcal{C}}}).
    \end{equation}
    Hence, given
    {the distribution $\CD$ of}
    $\mathbf{h}$, minimizing the mean squared quantization error, as defined in \autoref{meancoverdef}, is also equivalent to maximizing the average beamforming gain. 
    The smaller the average quantization error is for a codebook $\mathcal{C}$, the larger its average beamforming gain becomes.
\end{remark}

An alternative baseline for the beamforming gain is EGT \cite{Love_2003_EGT}, where all the coordinates of the EGT beamforming vector have equal amplitude (the equal-gain constraint),
with optimization of the beamforming gain only over the $N_{\text{T}}$ phases of the beamforming vector. 
The EGT gain under the equal gain beamforming vector is given by \cite{Love_2003_EGT} 
\begin{align}
\label{eqn:EGT gain}
    \Gamma_{\textup{EGT}}(\mathbf{h}) &= \frac{\|\mathbf{h}\|_1^2}{N_{\text{T}}}. 
\end{align}
Note that, naturally, $\Gamma_{\textup{EGT}}(\mathbf{h}) \leq \Gamma_\textup{MRT} (\mathbf{h})$. 

\subsection{MIMO system with two receive antennas}
\label{sec:MIMO system model}

For systems with $N_{\text{R}} = 2$, prior work~\cite{Ryan_2009_psk} has shown that codebooks obtained by quantizing the optimal beamforming vector $\mathbf{f}^{\rm opt}$ maximize the effective channel gain. Hence, we closely analyze a MIMO system with two receive antennas and aim to propose a codebook design that effectively quantizes the $\mathbf{f}^{\rm opt}$. Such a maximum gain is called the MRT gain, denoted by $\Gamma_{\textup{MRT}}$. Note that, the MRT gain is achievable with those codebooks without equal-gain constraint such as quadrature amplitude modulation (QAM) codebooks~\cite{Ryan_2009_psk}. We focus, however, on the beamforming gain achievable by a class of codebooks with equal-gain constraint with a motivation to effectively quantize the $N_{\text{T}}$ phases of the channel. Further, for such a codebook design, the baseline to consider is EGT gain for MIMO system for which a closed-form expression does not exist~\cite{Love_2003_EGT}. We compute this value and compare it against the CP
codebook gain in~\autoref{sec:iter method}.

\section{CP Codebook Beamforming}
\label{sec:iter method}

In this section, we describe how $\CCP$ codes are used to construct precoding codebooks for limited-feedback beamforming. We first describe the codebook construction, feedback design, and storage complexity, then explain how beamforming vector selection can be formulated as a decoding problem in the subspace coding domain. Finally, we present theoretical bounds on the quantization error and beamforming gain of $\CCP$ codebooks.

\subsection{CP codebook construction and feedback design}
\label{sec:Codebook construction}

Consider a $\CCP$ codebook $\mathcal{C}$ over a prime field $\fp$, as defined in \autoref{def:cp}, 
consisting of codewords $\CCP(f)$ for $f \in \CF_{p}(k, p)$. Then the size of the codebook for a given dimension $k$ is $|\mathcal{C}| = p^k$. The parameter $n$ of the $\CCP$ codebook is set as $n = N_{\text{T}}$. Note that each $\CCP$ codeword corresponds to a beamforming vector whose entries lie on the unit circle, thereby satisfying the equal-gain (constant-envelope) transmission constraint. Thus, $\CCP$ codewords can be directly used as precoding vectors. 

The feedback rate is determined by the code dimension $k$, which controls the size of the $\CCP$ codebook and therefore the beamforming resolution. Increasing $k$ increases the number of codewords exponentially and improves the quantization resolution of the beamforming vector. The number of feedback bits, denoted by $B$, from the receiver to the transmitter, to specify which beamforming vector from the $\CCP$ codebook to use, is given by
\begin{align*}
    B &= \lceil \log_2|\mathcal{C}| \rceil
    = \lceil \log_2(p^k) \rceil 
    = \lceil k \log_2 p \rceil. 
\end{align*}
The number of feedback bits $B$ therefore scales linearly with the dimension $k$ of the code and logarithmically with the 
number of antennas as $ n \le p-1 $ (one can pick the smallest $p > n$ for the codebook design). 

As discussed in \autoref{sec:System model} for a general codebook $\mathcal{C}$, the problem of searching for $\mathbf{f^{\rm opt}_{\textup{CP}}}$ over the $\CCP$ code is equivalent to finding the closest $\CCP$ codeword, in chordal distance, to $\mathbf{f^{\rm opt}}$ associated with the channel $\mathbf{h}$. Hence, this problem becomes equivalent to the decoding problem in the subspace coding domain, i.e., finding the closest codeword, in chordal distance, to a given vector (as a one-dimensional subspace). 

\subsection{Storage complexity}

The storage complexity of $\CC$ is determined by the evaluation points $\alpha_{1}, \dots, \alpha_{n} \in \fp^{\times}$ and the character function $\chi$. 
Note that there are exactly $p - 1$ non-trivial characters $\chi$ of $\fp$, so that specifying $\chi$ requires $\lceil \log_{2} (p - 1) \rceil = O(\log p)$ bits.
Therefore, the total storage cost for the $\CCP$ codebook is $O(n \log p)$, or $O(n \log n)$ by Bertrand's postulate when $p$ is the smallest prime greater than $n$. 
In particular, this is in sharp contrast to numerically optimized Grassmannian codebooks, which must store all $|\CC| = p^{k}$ codewords explicitly, each of length $n$, resulting in storage that grows exponentially in the feedback bits $B$. 
The algebraic structure of $\CCP$ codes thus provides a significant storage advantage: the entire codebook is implicitly defined by $O(n \log n)$ bits of parameters.

Any individual codeword $\CCP(f)$ can be computed on-the-fly by evaluating the polynomial $f$ at the stored points and applying $\chi$. 
This on-demand generation also means that codebook search
can be carried out without materializing the full codebook in
memory. 
Structured alternatives such as Kerdock codes~\cite{Inoue_2009_kerdock}
and NTCQ~\cite{au2011trellis, choi2013noncoherent} require 
precomputed codeword tables or trellis structures whose
memory footprint grows with $N_{\text{T}}$, and
none simultaneously achieves near-optimal Grassmannian packing
with a constant envelope constraint. 
{To assess the packing quality of $\CCP$ codebooks against this storage-optimal design, we include the AP Grassmannian codebook of~\cite{dhillon2008constructing} as a numerical benchmark in~\autoref{sec:Simulation}, with codebook size matched to CP at each $k$.} 
The $\CCP$ codebook, by
contrast, achieves both properties with storage and encoding
complexity comparable to classical Reed-Solomon
codes~\cite{wicker1999reed}.

\subsection{ 
Theoretical bounds
}

\label{sec:Covering radius}


To analyze the beamforming performance of $\CCP$ codebooks, we relate the quantization error to the covering radius of the corresponding Grassmann code. The covering radius determines the worst-case chordal distance between any beamforming direction and the nearest codeword, and therefore directly impacts the beamforming gain loss due to limited feedback. In this section 
we characterize and provide bounds on the quantization error of the $\CCP$ codebook and leverage these results to provide bounds on its beamforming gain. 

The following lemma provides a bound on the covering radius of $\GRS_{n}(\fpkq')$ and will be used later. 
\begin{lemma}
    \label{lem:rho-cp}
    The covering radius of $\GRS_{n}(\fpkq')$ satisfies 
    \begin{align}
        n - k 
        &\le \rhoh(\GRS_{n}(\fpkq')) 
        \le n - k + \bigg \lfloor \frac{k}{p} \bigg \rfloor. 
        \label{eq:grs-rho}
    \end{align}
    In particular, \hide{when $q$ is prime, }$\rhoh(\GRS_{n}(\fpkp')) = n - k$. 
\end{lemma}


\begin{proof}
    Let $\CC' = \GRS_{n}(\fpkq')$. 
    \hide{We claim that 
    \begin{align}
        n - k 
        &\le \rhoh(\CC') 
        \le n - k + \left \lfloor \frac{k}{p} \right \rfloor. 
        \label{eq:grs-rho}
    \end{align}
    To see this, o}Since $\CC' \subseteq \GRS$, we have $\rhoh(\CC') \ge \rhoh(\GRS) = n - k$. The right half of \eqref{eq:grs-rho} follows from 
    \eqref{eq:CPsize} and the redundancy bound
    ~\cite[Corollary~11.1.3]{Huffman03}. 
    %
\end{proof}

Let ${\bf y} = (y_1,\dots,y_n) \in \BC^{n}$ denote the input to the beamforming search/quantization problem. In the context of our problem we will have $
{\mathbf{y} = (\mathbf{f^{\rm opt}})^{\text{T}}}$ computed from the channel ${\mathbf{h}}$. 

Let $\Psi = \chi(\fq)$. Then $\Psi$ consists of the $p$-th roots of unity. Let also $\theta_{j} = \arg(y_{j}) \in [0, 2 \pi)$ 
and define 
\begin{align}
    \label{eq:ycheck}
    \check{y}_{j} 
    &:= \exp\bigg(\frac{2 \pi i}{p} \bigg \lfloor \frac{p \theta_{j}}{2 \pi} + \frac{1}{2} \bigg \rfloor\bigg) \hide{\\ 
    \tilde{y}_{j} 
    &:= y_{j} / |y_{j}| 
    = e^{i \theta_{j}}}
\end{align}
for $j \in \{1, \dots, n\}$. 
In other words, $\check{y}_{j}$ 
is the closest point 
to $y_{j}$ 
in $\Psi$ (in Euclidean distance)\hide{ and $\{z \in \BC: |z| = 1\}$, respectively for each $j$}. 

\hide{
\begin{align*}
    \theta 
    &:= \arccos\left\langle \frac{1}{\sqrt{n}} {\bf y}, \frac{1}{\sqrt{n}} \CCP(f) \right\rangle.
\end{align*}

\subsubsection{
Estimating the Quantization Error}

\begin{corollary}
    Let $\CCP = \CCP_{n}(\fpkp, \chi)$\hide{ and ${\bf y} \in \BC^{n}$ with $\norm{y} = \sqrt{n}$}. 
    Then chordal distance between any one-dimensional subspace in $\BC^{n}$ and its nearest $\CCP$ codeword is bounded above by 
    \begin{align}
        \sqrt{1 - \left(\frac{2 k}{n} - 1 - 2 \sin \frac{\pi}{2 p}\right)^{2}}. 
    \end{align}
    whenever 
    \begin{align*}
        k 
        &\ge \frac{n}{2} \left(2 \sin \frac{\pi}{2 p} + 1\right). 
    \end{align*}
\end{corollary}

\begin{proof}
    Note that 
    \begin{align*}
        D({\bf y}) 
        &:= \frac{1}{n} \norm{{\bf \check{y}} - {\bf \tilde{y}}}^{2} 
        = \frac{1}{n} \sum_{j = 1}^{n} 
        |e^{i \delta_{j}} - 1|^{2} \\ 
        &= \frac{1}{n} \sum_{j = 1}^{n} 
        [(1 - \cos \delta_{j})^{2} + \sin^{2} \delta_{j}] \\ 
        &=  \frac{1}{n} \sum_{j = 1}^{n} 
        4 \sin^{2} \frac{\delta_{j}}{2} \\ 
        &\le 4 \sin^{2} \frac{\pi}{2 p} 
        ,
    \end{align*}
    and the conclusion follows. 
\end{proof}
}

Note that the beamforming gain of a $\CCP$ code is given by 
\begin{align}
\label{eqn: CP gain theoretical}
\Gamma_{\textup{CP}}(\mathbf{h}) &= \frac{|\langle {\bf y}, \CCP(f)\rangle|^{2}}{n},     
\end{align}
where 
{${\bf y} = (\mathbf{f^{\rm opt}})^{\text{T}}$} associated with ${\mathbf{h}}$.

Also, recall that the EGT gain, which is the best beamforming gain with the equal-gain constraint under perfect CSI at the transmitter (CSIT), can be expressed 
as 
\begin{align}
\label{eqn: EGT gain norm 1}
    \Gamma_{\textup{EGT}}(\mathbf{h})&= \frac{\left(\sum_{i=1}^n |y_i|\right)^2}{n}.
\end{align} 
This motivates the following as a distortion measure of beamforming vector quantization compared with the baseline EGT. 
\begin{definition}[Normalized Distortion] 
\label{distortion-def}
    The \emph{normalized distortion} $\Delta_{\textup{CP}}$ for a given $\CCP$ code is defined as 
    \begin{align}
        \label{eq:qerror-new}
        \Delta_{\textup{CP}} 
        &:= \frac{\BE[\Gamma_{\textup{EGT}}({\mathbf{h}})] - \BE[\Gamma_{\textup{CP}}({\mathbf{h}})]}{\BE[\Gamma_{{\textup{EGT}}}({\mathbf{h}})]},
    \end{align}
    where the expectation is taken with respect to the distribution 
    {$\CD$} of ${\mathbf{h}}$.
\end{definition}
Note that $
{\mathbf{y} = (\mathbf{f}^{\rm opt})^{\text{T}} = \gamma \mathbf{h}^\dagger / \norm{\mathbf{h}}}$ is the unit-norm representative of the optimal beamforming direction, 
so that $|y_{i}| = |h_{i}|/\norm{{\bf h}}$, and \eqref{eqn: EGT gain norm 1} gives $\Gamma_{\rm EGT}({\bf h}) = \norm{{\bf h}}_{1}^{2} / (n \norm{{\bf h}}^{2})$, 
which differs from \eqref{eqn:EGT gain} by a factor of $\norm{{\bf h}}^{2}$. 
This is consistent since the normalized distortion $\Delta_{\CCP}$ in \autoref{distortion-def} is invariant to this scaling---the factor of $\norm{{\bf h}}^{2}$ cancels in the right-hand side of \eqref{eq:qerror-new}. 

In the next theorem, we upper bound the mean squared quantization error of the 
$\CCP$ codebook. Note that, as discussed in \autoref{rmk:qerror}, the quantization error relates to the beamforming gain of the $\CCP$ codebook. 

\hide{
Now, note that 
\begin{align*}
    \langle {\bf y}, {\bf \tilde{y}} \rangle 
    &= \sum_{j = 1}^{n} 
    |y_{j}|. 
\end{align*}
Our goal is to show that $\langle {\bf y}, \CCP(f) \rangle$ is close to $\langle {\bf y}, {\bf \tilde{y}} \rangle$. 
}

\begin{theorem}
    \label{thm:gain}
    Let $\CD$ be a distribution on the complex unit $n$-sphere\hide{ in $\BC^{n}$\hl{$\CS$}}, where the magnitudes are i.i.d. with mean $\mu$ and variance $\sigma^{2}$, and the magnitude and phase of each coordinate are independent. 
    Then the mean squared quantization error (\autoref{meancoverdef}) of a $\CCP$ code over $\fp$ of length $n$ and rate $R \ge 1 / (1 + \sqrt{\cos(2 \pi / p)})$ 
    is bounded as follows: 
    \hide{\begin{align}
         \left(\frac{1}{\sqrt{\cos(2 \pi / p)}} - 1\right)\frac{\sigma^{2}}{2 n \mu^{2}} <1,
        \label{eq:nbound}
    \end{align}
    ,}
    \begin{align}
        \label{eq:mean-covering-distance}
        \rhosbar(\CCP) 
        &\le 1 - 
        n \mu^{2} (R \sqrt{\cos(2 \pi / p)} + R - 1)^{2}
        . 
        \hide{\frac{2 \sigma^{2}}{\mu^{2} + \sigma^{2}}}
        \hide{&\le 2 \left[1 - \left(\frac{\mu^{2} + \sigma^{2} / n}{\mu^{2} + \sigma^{2}}\right) \cos\left(\frac{2 \pi}{p}\right)\right] \hide{\frac{2 \sigma^{2}}{\mu^{2} + \sigma^{2}}}
        2 \left(1 - \frac{c^{2} \mu^{2} + c^{2} \sigma^{2} / n}{\mu^{2} + \sigma^{2}}\right) \\ 
        &\hspace{-20pt} 2 \left(1 - \frac{(R (c + 1) - 1)^{2} \mu^{2} + \left(R (c^{2} - 1) + 1\right) \sigma^{2} / n}{\mu^{2} + \sigma^{2}}\right) 
        \notag }
    \end{align} 
    \hide{ and 
    \begin{align}
        R
        &\ge \frac{1}{c + 1} + \frac{(1 - c) \sigma^{2}}{2 (c + 1) n \mu^{2}} 
        ,
        \label{eq:rate-bound}
    \end{align}
    which approaches 
    In particular, 
    the \hide{asymptotic }mean squared quantization error as $n \to \infty$ is at most 
    \begin{align}
        \label{eq:asymptotic-mean-covering-distance}
        2 \left(1 - \frac{(2 R - 1)^{2} \mu^{2}}{\mu^{2} + \sigma^{2}}\right),
    \end{align}
    as $n \to \infty$, }
\end{theorem}

\begin{proof}
    Let $\CCP = \CCP_{n}(\fpkp, \chi)$ and ${\bf y} \in \BC^{n}$ with $\norm{y} = 1$. Observe that 
    \begin{align}
        |\langle {\bf y}, \CCP(f) \rangle| 
        &\ge 
        \abs{\abs{\sum_{i \in A} y_{i}^{*} \chi(f(\alpha_{i}))} - \abs{\sum_{i \in B} y_{i}^{*} \chi(f(\alpha_{i}))}} \notag \\ 
        &\ge \abs{\sum_{i \in A} y_{i}^{*} \chi(f(\alpha_{i}))} - \sum_{i \in B} |y_{i}| 
        \label{eq:inner-product}
    \end{align}
    almost surely by the triangle inequality, 
    where 
    \begin{align*}
        \begin{aligned}
            A &= \{1 \le i \le n: \check{y}_{i} = \chi(f(\alpha_{i}))\}
            \\ 
            B &= \{1 \le i \le n: \check{y}_{i} \neq \chi(f(\alpha_{i}))\}
        \end{aligned}
    \end{align*}
    so that \hide{
    \begin{align*}
        \abs{\sum_{i \in A} \check{y}_{i}^{*} \chi(f(\alpha_{i}))} 
        &= |A| 
    \end{align*} 
    and }$|B| \le 
    n - k$ by \autoref{lem:rho-cp} combined with the fact that $\chi: \fp \to \BC$ is injective. 
    
    To estimate the first term, note that by \eqref{eq:ycheck}, $\check{y}_{j} = e^{i \delta_{j}} y_{j} / {|y_{j}|}$ 
    for $j \in \{1, \dots, n\}$, 
    where 
    \begin{align}
        \label{eq:delta-j}
        \delta_{j} 
        &:= \frac{2 \pi}{p} \left \lfloor \frac{p \theta_{j}}{2 \pi} + \frac{1}{2} \right \rfloor - \theta_{j} 
        \in \left[- \frac{\pi}{p}, \frac{\pi}{p}\right),
    \end{align}
    so that 
    \begin{align}
        \label{eq:y-delta-j}
        y_{j}^{*} \chi(f(\alpha_{j})) 
        &= y_{j}^{*} \check{y}_{j} 
        = |y_{j}| e^{i \delta_{j}}. 
    \end{align}
    Thus, 
    \begin{align}
        &\hspace{-2em} \abs{\sum_{j \in A} y_{j}^{*} \chi(f(\alpha_{j}))}  
        = \abs{\sum_{j \in A} |y_{j}| e^{i \delta_{j}}} 
        \notag \\ 
        &= \sqrt{\sum_{j \in A} |y_{j}|^{2} + 2 \sum_{\substack{j, \ell \in A \\ j < \ell}} |y_{j}| |y_{\ell}| \cos (\delta_{j} - \delta_{\ell})} 
        \notag \\ 
        &\ge \sqrt{\sum_{j \in A} |y_{j}|^{2} + 2 \sum_{\substack{j, \ell \in A \\ j < \ell}} |y_{j}| |y_{\ell}| \cos (2 \pi / p)} 
        \notag \\ 
        &\ge 
        \sqrt{\cos(2 \pi / p)}
        \sum_{j \in A} |y_{j}| 
        \label{eq:pathwise}
    \end{align}
    by \eqref{eq:delta-j} and \eqref{eq:y-delta-j}. 
    Therefore, taking expectations in \eqref{eq:inner-product} and using \eqref{eq:pathwise} together with magnitude-phase independence,
    \hide{
    \begin{align}
        \langle {\bf y}, \CCP(f) \rangle 
        &\ge 
        \sqrt{\cos(2 \pi / p)}
        \sum_{j \in A} 
        |y_{j}| - 
        \sum_{j \in B} |y_{j}| 
        \label{eq:ABbound}
        ,
    \end{align}
    i.e. }
    \begin{align}
        \BE[|\langle {\bf y}, \CCP(f) \rangle|] \ge (c a - b) \mu 
        &\ge (c k + k - n) \mu,
        \label{eq:exp-inner-product}
    \end{align} 
    where $a = |A|$, $b = |B| = n - a \le n - k$ and $c = \sqrt{\cos(2 \pi / p)}$. 
    \hide{Now suppose that the right-hand side of \eqref{eq:ABbound} is negative. Switching $A$ and $B$ in the argument above gives
    \begin{align*}
        \langle {\bf y}, \CCP(f) \rangle 
        &\ge 
        \abs{\sum_{i \in B} y_{i}^{*} \chi(f(\alpha_{i}))} - 
        \sum_{j \in A} |y_{j}| \\ 
        &\ge 
        \abs{\sum_{i \in B} y_{i}^{*} \chi(f(\alpha_{i}))} - \frac{1}{\sqrt{\cos(2 \pi / p)}} 
        \sum_{j \in B} |y_{j}| 
    \end{align*}
    }Thus, assuming $R = k / n \ge 1 / (c + 1)$, we have 
    \begin{align}
        \BE[|\langle {\bf y}, \CCP(f) \rangle|^{2}] 
        &\ge \BE[|\langle {\bf y}, \CCP(f) \rangle|]^{2} \ge 
        (c k + k - n)^{2} \mu^{2} 
        \label{eq:exp-inner-product-squared}
    \end{align}
    \hide{Assuming the $|y_{j}|$ are i.i.d.,\hide{ with mean $\mu$ and variance $\sigma^{2}$} the probability that the right-hand side of \eqref{eq:ABbound} is negative is 
    \begin{align}
        \hide{\Pr[{\rm RHS} < 0] 
        &= }\Pr\left[\frac{\sum_{j \in B} |y_{j}|}{\sum_{j \in A} |y_{j}|} > c\right]  
        \le \frac{b}{c a} 
        \le \frac{n - k}{c k} = \frac{1 - R}{c R}
        \label{eq:markov}
    \end{align}
    by Markov's inequality, 
    Therefore, \hl{as $R\to 1$},\hide{ which tends to $0$ as $R \to 1$, i.e. ${\rm RHS} \ge 0$ almost surely$\BE[{\rm RHS}] = (c a - b) \mu \ge (c k + k - n) \mu > 0$ so the RHS is positive on average! In particular, (This is only true if the RHS of \eqref{eq:ABbound} is positive! Q: If $X \ge Y$ and $\BE[Y] \ge 0$, is $\BE[X^{2}] \ge \BE[Y^{2}]$?
    
    Note: iff $\var(X) + \BE[X]^{2} \ge \var(Y) + \BE[Y]^{2}$. In the worst case, we cannot say anything more than this since ${\bf y}$ can be perpendicular to $\CCP(f)$. Nevertheless, $\BE[{\rm RHS}] = (c a - b) \mu \ge (c k + k - n) \mu > 0$ so the RHS is positive on average! In particular, by Markov's inequality, (see \href{https://math.stackexchange.com/questions/1258485/expectation-of-quotient-of-random-variables}{here})
    \begin{align*}
        \Pr[{\rm RHS} < 0] 
        &= \Pr\left[\frac{\sum_{j \in B} |y_{j}|}{\sum_{j \in A} |y_{j}|} > c\right]  
        \le \frac{b}{c a} 
        \le 
        \frac{1 - R}{c}
    \end{align*}
    which tends to $0$ as $R \to 1$, i.e. ${\rm RHS} \ge 0$ almost surely!)} 
    \begin{align}
        \expec
        [\langle {\bf y},  \CCP(f) \rangle^{2}] 
        &\ge \expec
        \bigg[\bigg(\sqrt{\cos(2 \pi / p)} 
        \sum_{j \in A} |y_{j}| - \sum_{j \in B} |y_{j}|\bigg)^{2}\bigg] \notag \\ 
        &= (ca - n + a)^{2} \mu^{2} + (c^{2} a + n - a) \sigma^{2}
        \label{eq:g(a)}
    \end{align}
    \hl{almost surely}. 
    \eqref{eq:g(a)} is \hide{strictly convex and }increasing in $a$ for $a \ge n R = k$ 
    and 
    \begin{align}
        R
        &\ge \frac{1}{c + 1} + \frac{(1 - c) \sigma^{2}}{2 (c + 1) n \mu^{2}} 
        . 
        \label{eq:rate-bound}
    \end{align}
    Assuming \eqref{eq:nbound} holds, \eqref{eq:rate-bound} is satisfied for $R \to 1$. 
    Therefore, 
    \begin{align*}
        \expec
        [\langle {\bf y}, \CCP(f) \rangle^{2}]
        &\ge 
        (c k + k - n)^{2} \mu^{2} + (c^{2} k + n - k) \sigma^{2}
        ,
    \end{align*}
    }by the Cauchy--Schwarz inequality and \eqref{eq:exp-inner-product}. Therefore, 
    \begin{align*}
        \rhosbar(\CCP)
        &= 1 - \frac{\BE[|\langle {\bf y}, \CCP(f) \rangle|^{2}]}{n} 
        \le 
        1 - \frac{(c k + k - n)^{2} \mu^{2}\hide{ + (c^{2} k + n - k) \sigma^{2}}}{n}, 
    \end{align*}
    as desired. 
\end{proof}

\hide{As $R \to 1$, the RHS $\to 1 - c^{2} \mu^{2} - c^{2} \sigma^{2} / n$.  

\eqref{eq:rhostilde} assumes that $\norm{{\bf y}} = \sqrt{n}$, so adding this constraint to the hypothesis of \autoref{thm:gain} yields a clean bound. Note that this does not affect $\langle {\bf y} \rangle$, and we can replace $\sqrt{n}$ with any constant (e.g. $1$, which would instead require $\mu^{2} + \sigma^{2} = 1 / n$ and introduce a factor of $n$ on the RHS\hide{ of \eqref{eq:rhostilde}}) depending on the hypothesis on ${\mathbf{h}}$. 

\hide{(Addendum) We can also get rid of the constraint altogether: }
\begin{align}
    \hide{\frac{\rhos(\CCP)}{2} 
    &\ge }
    \frac{\rhosbar(\CCP)}{2} 
    &\le 
    1 - \frac{(c k - k + n)^{2} \mu^{2} + (c^{2} k + n - k) \sigma^{2}}{n^{2}} 
    \label{eq:rhostilde}\\ 
    &= 1 - (R (c + 1) - 1)^{2} \mu^{2} - \left(\frac{R (c^{2}  - 1) + 1}{n}\right) \sigma^{2}
    \notag 
\end{align}
Note that computing the exact value of the expectation requires knowledge of the distribution of $\norm{{\bf y}}$. 
\hide{In general, 
\begin{align*}
    f_{\norm{{\bf y}}}(t) d t 
    &= \frac{d}{dr} \Pr[\norm{{\bf y}} \le r] 
    = \frac{d}{dr} \Pr\left[\sum_{i = 1}^{n} |y_{i}|^{2} \le r^{2}\right], 
\end{align*}
which requires knowledge of the distribution of $y_{i}$. }
}

The next theorem shows that the normalized distortion of the $\CCP$ codebook, defined in \autoref{distortion-def}, approaches $0$ as the length $n=N_{\text{T}}$ grows large and the code rate approaches $1$.

\begin{theorem}
    \label{thm:cp-gain}
    Let \hide{$\CCP = \CCP_{n}(\fpkp, \chi)$ and }$\CD$ be a distribution on $\BC^{n}$, where the magnitudes are i.i.d. with mean $\mu > 0$ and variance $\sigma^{2}$, and the magnitude and phase of each coordinate are independent. 
    Then, for a $\CCP$ code over $\fp$ of length $n$ and rate $R \ge 1 / (1 + \sqrt{\cos(2 \pi / p)})$,\hide{
    \begin{align*}
        \Delta_{\CCP} 
        &\le 1 - \frac{(R (c + 1) - 1)^{2}}{1 + \sigma^{2} / (n \mu^{2})},
    \end{align*}
    where the $\CCP$ code rate $R \ge 1 / (c + 1)$ and $c = \sqrt{\cos(2 \pi / p)}$. }
    \begin{align}
        \label{eq:quant-error}
        \Delta_{\CCP} 
        &\le 1 - \frac{(R \sqrt{\cos(2 \pi / p)} + R - 1)^{2}}{1 + \sigma^{2} / (n \mu^{2})}. 
    \end{align}
    In particular, $\Delta_{\CCP} \to 0$ as $n \to \infty$ and $R \to 1$.
    \hide{for a $\CCP$ code of rate $R \to 1$
    , }
    \hide{
    \begin{align*}
        \lim_{n \to \infty} \frac{\BE[\Gamma_{\textup{CP}}({\mathbf{h}})]}{\BE[\Gamma_{{\textup{EGT}}}({\mathbf{h}})]} 
        &\ge (2 R - 1)^{2}.
    \end{align*}
    the ratio of the mean beamforming and mean EGT gains is at least 
    \begin{align}
        & (R (c + 1) - 1)^{2} + \left(\frac{R (c^{2}  - 1) + 1}{n}\right) \left(\frac{\sigma}{\mu}\right)^{2}.
    \end{align}
    In particular, the quantization error
    \begin{align}
        \frac{\BE[\Gamma_{\textup{CP}}({\mathbf{h}})]}{\BE[\Gamma_{{\textup{EGT}}}({\mathbf{h}})]} 
        &\ge \cos\left(\frac{2 \pi}{p}\right),
    \end{align}
    i.e.
    \begin{align}
        \label{eq:n-limit} 
        \lim_{n \to \infty} \Delta_{\textup{CP}} &= 0. 
    \end{align} } 
    \hide{\begin{align*}
        \lim_{R \to 1} 
        \lim_{n \to \infty} \frac{\BE[\Gamma_{\textup{CP}}({\mathbf{h}})]}{\BE[\Gamma_{{\textup{EGT}}}({\mathbf{h}})]} 
        &= 1.
    \end{align*}}
\end{theorem}

\begin{proof}
    Using the notation and following the derivations in the proof of \autoref{thm:gain}, we have 
    \begin{align}
        \label{eq:gain-rho}
        \frac{\BE\left[
        \Gamma_{\textup{CP}}({\mathbf{h}})
        \right]}{\BE[\Gamma_{{\textup{EGT}}}({\mathbf{h}})]} 
        &\ge \frac{(c k + k - n)^{2} \mu^{2}}{n^{2} \mu^{2} + n \sigma^{2}}
        = \frac{(R (c + 1) - 1)^{2}}{1 + \sigma^{2} / (n \mu^{2})} 
    \end{align}
    by \eqref{eq:exp-inner-product-squared} and \eqref{eqn: EGT gain norm 1}. Now \eqref{eq:quant-error} follows from \eqref{eq:qerror-new} and \eqref{eq:gain-rho}, while the final claim follows from \eqref{eq:quant-error} upon noting that $n < p$. 
\end{proof}
  ~\autoref{thm:gain} and~\autoref{thm:cp-gain} characterize the quantization error and normalized distortion of $\CCP$ codebooks for MISO systems, where the EGT baseline admits the closed-form expression in~\eqref{eqn:EGT gain}. 
  {These bounds assume $\CD$ to have i.i.d. coordinate magnitudes with independent phase. Channel models without this structure are evaluated empirically.} 

For MIMO systems with multiple receive antennas, no such closed-form expression exists~\cite{Love_2003_EGT}, and the EGT baseline must be computed numerically. The problem of establishing this baseline is non-convex due to the equal-gain constraint on the beamforming vectors. We propose an iterative approach to compute the EGT baseline for the MIMO system, which is then used to evaluate the $\CCP$ codebook performance.


\subsection{EGT gain for MIMO system with \texorpdfstring{$N_{\text{R}} = 2$}{NR = 2}}
\label{sec:EGT for Nr_2}
The gain of EGT~\eqref{eqn:EGT gain} establishes the baseline that any equal-gain codebook design for the MISO system should approach while maintaining minimal storage complexity. For a MIMO system with two receiver antennas, the characterization of this baseline is non-trivial. Let us closely examine the beamforming gain maximization problem by rewriting~\eqref{eqn:beamforming gain} for codebooks that satisfy the equal-gain constraint.

\begin{align}
    &\max_{\mathbf{f \in \mathcal{C}}}{\|\mathbf{Hf}\|^2}\nonumber\\
    &\text{s.t. } |f_i| = 1/\sqrt{N_{\text{T}}} \quad \forall i \in \{1,\ldots,N_{\text{T}}\}\label{eqn:EGT characterization constraint}
\end{align}
Note that the maximization above is written by noting $\mathbf{z} = \mathbf{Hf}/\|\mathbf{Hf}\|$. The EGT maximization for MIMO systems now becomes a non-convex optimization problem since the set of constraints in~\eqref{eqn:EGT characterization constraint} are along the points on unit complex circle. This problem is identified in~\cite{Love_2003_EGT} for a set of general equal gain transmission and receiver systems. The authors propose a quantized version of EGT gain that does not compromise on the diversity of the MIMO system but offers a tradeoff between the number of bits used to decide the levels of phase quantization and the complexity of the codebook search. Prior work~\cite{Murthy_EGT} proposes both vector and scalar quantization techniques to quantize the
phase angles and establish the quantized EGT. A transmitter antenna selection algorithm~\cite{Tsai_EGT} reduces the SNR gap between the MRT
gain for a small number of antennas. The established results also characterize the capacity loss and probability of outage. 


In this paper, we use an iterative approach. In~\autoref{alg:Iterative EGT update}, for simplicity, we consider the beamforming gain as $\Gamma$ which is a function of the underlying channel $\mathbf{H}$. In Step 3, we initialize the beamforming vector as the normalized version of the $\mathbf{f}^{\rm opt}$ preserving the angles given by the channel. 
The algorithm proceeds to iteratively update all $N_{\text{T}}$ phases for any $b$-bit quantization. At Step 9, we initialize the $k^{\rm th}$ phase of $\mathbf{\tilde{f}}$ with the phase given by $\mathbf{f}^{\rm opt}$, while the remaining $(N_{\text{T}} - k)$ phases remain unchanged, which are yet to be optimized. From Steps 12 to 19, the $k^{\rm th}$ phase is determined and the algorithm now proceeds to update the $(k+1)^{\rm th}$ phase, where the remaining $N_{\text{T}} - (k + 1)$ phases are yet to be updated. When the algorithm terminates, $\Gamma^{(0)}$ will store the value of $\Gamma_{\textup{EGT}}(\mathbf{H})$. Note that the algorithm will run the $N_{\rm sim}$ Monte Carlo simulations with different realizations of $\mathbf{H}$ and $\mathbf{f}^{\rm opt}$ to compute the EGT gain. 

\begin{algorithm}
    \caption{\textit{Iterative EGT update}}
    \begin{algorithmic}[1] 
        \State {\bf Input:} MIMO channel ${\bf H} \in \BC^{N_{\text{R}} \times N_{\text{T}}}$, ${\bf f}^{{\rm opt}} \in \BC^{N_{\text{T}} \times 1}$, number of iterations $N \in \mathbb{N}$
        \State {\bf Output:} EGT beamforming baseline gain $\Gamma_{{\textup{EGT}}}({\bf H})$
        \State $\mathbf{f}^{(0)} \gets \frac{1}{\sqrt{N_{\text{T}}}}\mathbf{f}^{\rm opt}$.
        \State $\tilde{\bf f} \gets {\bf f}^{(0)}$
        \State $\Gamma^{(0)} \gets \|\mathbf{H \tilde{f}}\|^2 $ 
        \For{$\ell \in \{1, \dots, N\}$}
            \For{$m \in \{1, \dots, N_{\text{T}}\}$}
                \For{$i \in \{m + 1, \dots, N_{\text{T}}\}$}
                    \State $\tilde{f}_{i} \gets \frac{1}{\sqrt{N_{\text{T}}}} e^{j \arg(f_{i}^{(0)})}$
                \EndFor
                \State ${\bf \tilde{f}} \gets [f_{1}^{(0)}, \dots, f_{m}^{(0)}, {\bf \tilde{f}}]$
                \For{$w \in \{0, \dots, 2^b-1\}$}
                    \State $\phi \leftarrow \frac{1}{\sqrt{N_{\text{T}}}}e^{j2\pi w/2^b}$
                    \State $\hat{\mathbf{f}}^{(w)} \leftarrow [\tilde{f}_1, \ldots, \tilde{f}_{m-1}, \phi, \tilde{f}_{m+1}, \ldots, \tilde{f}_{N_{\text{T}}}]$
                    \State $\Gamma_{m}^{(w)} \gets \norm{{\bf H} \tilde{\bf f}^{(w)}}^{2}$
                    \If{$\Gamma^{(w)}_{m} > \Gamma^{(0)}$}
                        \State $\Gamma^{(0)} \gets \Gamma^{(w)}_{m}$
                        \State $f_{m}^{(0)} \gets \frac{1}{\sqrt{N_{\text{T}}}}e^{j2\pi w/{2^b}}$
                    \EndIf
                \EndFor
            \EndFor
        \EndFor
        \State \Return $\Gamma^{(0)}$
    \end{algorithmic}
    \label{alg:Iterative EGT update}
\end{algorithm}

At the same time, we observe that there is a tradeoff between $b$ and $N$, the parameters that govern the speed with which the algorithm converges to the true EGT baseline gain depending on the channel conditions $\mathbf{H}$. In the next subsection we extend our MISO and MIMO $\CCP$ codebook design to correlated channels.

\subsection{CP codebook design for correlated channel conditions}
\label{sec: CP dsign for correlated channels}

Antenna spatial correlation changes both the channel capacity and the degrees of freedom, which in turn affect the codebook design. Dense antenna arrays in emerging 6G systems produce spatially correlated channels~\cite{Guo_6G_antenna,liu2024nearfield}. Further, there exists stochastic modeling of the channel correlation with its effect on the multiplexing gains~\cite{Tse_Viswanath_2005}. Prior work~\cite{love2004grassmannian} incorporated transmit and receive antenna correlation into the channel
model for beamforming codebook design. In our system model for MISO systems, the correlated channel between the transmitter and the receiver is now represented as 
\begin{align}
\label{eqn:corr MISO channel}
 {\tilde{\mathbf{h}}^{\text{T}}} = \mathbf{R}_{\text{RX}} 
 {\mathbf{h}^{\text{T}}} \mathbf{R}_{\text{TX}},   
\end{align}
where the transmitter correlation matrix is $ \mathbf{R}_{\text{T}} = \mathbf{R}_{\text{TX}}\mathbf{R}_{\text{TX}}^\dagger$, the receiver correlation matrix is $\mathbf{R}_\text{R} = \mathbf{R}_{\text{RX}}\mathbf{R}_{\text{RX}}^\dagger$ and $\mathbf{h}$ is the uncorrelated channel. Note that for the MISO system, $\mathbf{R}_{\text{RX}} = 1$.  

The matrix $\mathbf{R}_{\text{TX}}$ is generated by the Cholesky decomposition of the correlation matrix $\mathbf{R}_\text{T}$. The correlation matrix $\mathbf{R}_\text{T}$ follows an exponential distribution with a known correlation coefficient between antenna elements. The approach to generate the correlation matrices is a standard technique proposed in~\cite{Kermoal_corr, Shiu_corr}. Thus, following a similar approach to generate $\mathbf{R}_{\text{RX}}$ using different correlation coefficients to reflect the different environments between the transmitter and receiver, 
we represent the correlated MIMO channel as
\begin{align}
\label{eqn:corr MIMO channel}
    \tilde{\mathbf{H}} = \mathbf{R}_{\text{RX}} \mathbf{H} \mathbf{R}_{\text{TX}},
\end{align}
where $\mathbf{H}$ is the uncorrelated fading MIMO channel. The question of how to design beamforming codebooks under this correlated channel model has been addressed in prior work through codebook rotation~\cite{love2004grassmannian}, but this approach has important implications for equal-gain codebook designs.




\begin{remark}
    When this rotation approach is applied to equal-gain codewords, the constant-envelope property no longer holds since the individual antenna entries no longer have equal magnitude. As a result, the EGT baseline is no longer a valid performance benchmark for the rotated codebook. In our CP codebook design, the codewords retain the equal-gain property, i.e., we do not rotate the codewords and continue to search the CP codebook without any rotation for correlated channels, and the codebook achieves EGT baseline performance, as observed in the next section.
\end{remark}

\section{Simulations}
\label{sec:Simulation}

{
In this section, we present simulation results for the beamforming gain 
of the proposed precoding design using 
$\CCP$ codebooks for MISO systems~\cite{gooty2025precodingdesignlimitedfeedbackmiso} and MIMO systems with two receiver antennas for different types of channel profiles. The simulations focus on rank-1 (single-stream) precoding, which is the primary scope of the proposed $\CCP$ codebook framework for MISO and MIMO systems. 

In particular, we consider the standard Rayleigh fading channel which models the rich scattering environments along with other correlated channels such as exponentially correlated channels under low and high correlated environments. To further evaluate the robustness of the $\CCP$ codebook design, we consider standardized CDL channel models under a limited feedback setup. The simulation results are presented as beamforming gain versus feedback bits $B$, where increasing $B$ corresponds to finer quantization of the beamforming direction. For $\CCP$ codebooks, this is controlled by the code dimension $k$, which determines the size of the codebook. Thus, increasing $k$ improves the Grassmannian packing resolution and allows the codebook to better approximate the optimal beamforming direction.

The baseline comparison is $\Gamma_{\textup{EGT}}(\mathbf{h})$ in \eqref{eqn:EGT gain} with perfect CSIT
for the MISO system.  
For the MIMO system with $N_{\text{R}}=2$, we consider the baseline EGT,  $\Gamma_{\textup{EGT}}(\mathbf{H})$ calculated by~\autoref{alg:Iterative EGT update}. 
Also, note that $N_{\text{T}} = n = p-1$.

\subsection{Feedback bits across codebooks}
To ensure a fair comparison across codebook designs, we match the number of feedback bits $B$ as closely as possible for each operating point. For the $\CCP$ codebook, $B = \lceil k \log_2 p \rceil$ is determined by the code dimension~$k$. For the DFT codebook, we select the oversampling factor $O_1$ such that $\lceil \log_2(N_{\text{T}} \cdot O_1) \rceil$ is closest to the target~$B$. For the PSK codebook, we choose the constellation order~$M$ so that $\lceil(N_{\text{T}} - 1)\log_2 M\rceil$ matches the target~$B$. Exact matching is not always possible due to the discrete structure of each codebook family, and where mismatches occur, we report the codebook at the nearest achievable feedback rate.

}

\subsection{MISO system}
\subsubsection{Rayleigh Fading}
\label{sec:Rayleigh fading}
The time-domain channel $\mathbf{h}$ is a memoryless complex Gaussian channel with unit variance. 
The simulations were carried out in a SageMath (v10.3) environment by averaging the gains with 300 Monte Carlo simulations. 
As observed in \autoref{Res: Rayleigh fading}, $\Gamma_{\textup{CP}}(\mathbf{h})$ is initially low for lower rate $\CCP$ codes. With 
$k$ approaching $n$, however, $\Gamma_{\textup{CP}}(\mathbf{h})$ approaches $\Gamma_{\textup{EGT}}(\mathbf{h})$, 
demonstrating a smooth trade-off between the number of feedback bits 
and the $\CCP$ beamforming gain. This behavior can be explained by the fact that increasing $k$ increases the size of the $\CCP$ codebook exponentially, resulting in finer Grassmannian packing and reduced quantization error. Consequently, the selected beamforming vector aligns more closely with the optimal equal-gain beamforming vector. 
Also, the $\CCP$ gain approaching the EGT gain agrees with the conclusion of \autoref{thm:cp-gain}. 
{To quantify the packing penalty of the algebraic $\CCP$ codebook structure, we include 
$\Gamma_{\text{AP}}(\mathbf{h})$ for $N_{\text{T}} = 4$, the gain of a numerically optimized 
Grassmannian codebook constructed via AP~\cite{dhillon2008constructing}, 
with codebook size matched to $\CCP$ codebook at each $k \leq 3$ (runtime prohibits $k=4$ at 
$p^k = 625$ codewords). As seen in Figure~\ref{Res: Rayleigh fading}, the AP curve lies 
within $0.15$~dB of $\Gamma_{\text{CP}}(\mathbf{h})$ across all operating 
points, confirming that the structured $\CCP$ codebook incurs negligible packing loss.}

\begin{figure}[!htbp]
\centering
  {\begin{tikzpicture}
\definecolor{black}{rgb}{0, 0.0, 0}%
\definecolor{blue}{rgb}{0, 0, 1}%
\definecolor{red}{rgb}{1, 0, 0}%

\definecolor{apcolor}{HTML}{7E2F8E}%


\begin{axis}[
font=\footnotesize,
width=0.65\columnwidth,
height=4.5cm,
scale only axis,
xmin=1,
xmax=6,
xtick = {1,2,3,...,6},
xlabel={CP code dimension ($k$)},  
xmajorgrids,
ymin= 3,
ymax= 10,
ytick = {3, 4, ..., 10},
ylabel={${{\Gamma_{\text{ave}}}}[\text{dB}]$},
ylabel near ticks,
ymajorgrids,
grid style={gray!80},
legend style={font=\small, at={(-0.125,1.03)},anchor=south west, draw=black,
fill=white,legend cell align=left},
legend columns = 3
]

\addplot [color=red, solid, densely dotted, mark=star, mark options = {solid}, line width=1.2pt]
  table[row sep=crcr]{%
1 5.143838021468881\\
2 5.143838021468881\\
3 5.143838021468881\\
4 5.143838021468881\\
};
\addlegendentry{$\Gamma_\text{EGT}(\mathbf{h_{ 4}})$}

\addplot [color=black, solid, densely dotted, mark=diamond, mark options = {solid}, line width=1.2pt]
  table[row sep=crcr]{%
1 6.923975494083980\\
2 6.923975494083980\\
3 6.923975494083980\\
4 6.923975494083980\\
5 6.923975494083980\\
6 6.923975494083980\\
};
\addlegendentry{$\Gamma_\text{EGT}(\mathbf{h_{ 6}})$};

\addplot [color=blue, solid, densely dotted, mark=triangle, mark options = {solid}, line width=1.2pt]
  table[row sep=crcr]{%
1 9.04229267371817\\
2 9.04229267371817\\
3 9.04229267371817\\
4 9.04229267371817\\
5 9.04229267371817\\
6 9.04229267371817\\
7 9.04229267371817\\
};
\addlegendentry{$\Gamma_\text{EGT}(\mathbf{h_{10}})$};

\addplot [color=red, dash dot, mark=asterisk, mark options = {solid}, line width=1.2pt]
  table[row sep=crcr]{%
1 3.395852139379703\\
2 4.463140368484376\\
3 4.897939326651402\\
4 4.897939327002356\\
};
\addlegendentry{$\Gamma_\text{CP}(\mathbf{h_{4}})$};

\addplot [color=black, dash dot, mark=diamond, mark options = {solid}, line width=1.2pt]
  table[row sep=crcr]{%
1 4.036782267057505\\
2 5.545254644279450\\
3 6.227571559567807\\
4 6.618367314149695\\
5 6.778818102780297\\
6 6.778818103171327\\
};
\addlegendentry{$\Gamma_\text{CP}(\mathbf{h_{6}})$};

\addplot [color=blue, solid, dash dot, mark=triangle, mark options = {solid}, line width=1.2pt]
  table[row sep=crcr]{%
1 4.616540793440172\\
2 6.598710734736112\\
3 7.588866867244284\\
4 8.157901154719834\\
5 8.504725197\\
6 8.725754903\\
};
\addlegendentry{$\Gamma_\text{CP}(\mathbf{h_{10}})$};

\addplot [color=apcolor, dashed, mark=o, mark options = {solid}, line width=1.5pt]
  table[row sep=crcr]{%
1 3.309038737676101\\
2 4.422727843437298\\
3 5.051340872172261\\
};
\addlegendentry{$\Gamma_\text{AP}(\mathbf{h_{ 4}})$};

\node[anchor=west] at (3.05,4.4){\textcolor{red}{$N_T = 4$}};
\node[anchor=north] at (4.5,6.6){\textcolor{black}{$N_T = 6$}};
\node[anchor=north] at (5.5,8.6){\textcolor{blue}{$N_T = 10$}};

\end{axis}

\end{tikzpicture}%
}
  \caption{Average beamforming gain $\Gamma_\text{ave}[\text{dB}]$ as a 
  function
  of the $\CCP$ code dimension $k$ for Rayleigh fading channel in 
    MISO system. The $\CCP$ codebook approaches the EGT baseline as $k$ grows, consistent with the bound in~\autoref{thm:cp-gain}. Three antenna configurations ($N_{\text{T}}=4,6,10$) show the convergence trend.} 
  \label{Res: Rayleigh fading}  
\end{figure}



\subsubsection{Correlated rayleigh fading}
The correlated channel is modeled using an exponential correlation matrix with transmit correlation coefficient $\rho_{\text{TX}}$, following~\eqref{eqn:corr MISO channel}. \autoref{fig:miso_rayleigh_exp} presents the average beamforming gain versus feedback bits $B$ for a 
MISO system ($N_{\text{T}} = 10$) with low correlation ($\rho_{\text{TX}} = 0.2$) and high correlation ($\rho_\text{TX} = 0.75$). In both regimes, $\CCP$ gain approaches EGT gain with increasing $B$. The $\CCP$ codebook achieves this performance without any codebook rotation or adaptation
based on channel statistics. This highlights a key advantage of $\CCP$ codebooks. Their Grassmannian structure provides robustness to channel correlation without requiring transmitter-side knowledge of the correlation matrix. In contrast, DFT saturates below the EGT baseline in the low correlation regime. Although it improves for lower values of $B$ under high correlation, this behavior is consistent with its design assumption of spatially correlated ULA channels. PSK remains largely insensitive to increasing feedback bits in both regimes, confirming the suboptimal Grassmannian packing of constellation-based codebooks.


\begin{figure*}[!htbp]
  \centering
  \hspace{0.75cm}
  \subfloat[Low correlation\label{fig:miso_rayleigh}]{%
    \begin{minipage}[t]{0.45\textwidth}
      \begin{tikzpicture}
\definecolor{black}{rgb}{0, 0.0, 0}%
\definecolor{blue}{rgb}{0, 0, 1}%
\definecolor{red}{rgb}{1, 0, 0}%



\begin{axis}[
font=\footnotesize,
width=0.7\columnwidth,
height=4.5cm,
scale only axis,
xmin=1,
xmax=25,
xtick = {0,5,10,...,25},
xlabel={Feedback bits ($B$)},  
xmajorgrids,
ymin= 4,
ymax= 10,
ytick = {4,5, ..., 10},
ylabel={$\Gamma_{\text{ave}}[\text{dB}]$},
ylabel near ticks,
ymajorgrids,
grid style={gray!80},
legend style={
  font=\small, 
  at={(0.5,1.02)}, 
  anchor=south, 
  draw=black, 
  fill=white, 
  legend cell align=center,
  /tikz/column 1/.append style={column sep=5pt},
  /tikz/column 2/.append style={column sep=5pt},
  /tikz/column 3/.append style={column sep=5pt},
},
legend columns=2,
]

\draw[->, thick, >=stealth] (axis cs:18,1.05) -- (axis cs:20,1.0);
\node[font=\scriptsize, anchor=east] at (axis cs:18,1.05) {EGT};


\addplot [color=black, line width=1.5pt, densely dotted, mark=o, mark size=1.5pt, mark options={solid, fill=blue}, mark repeat=1]
table[row sep=crcr]{%
0 9.004402557594164\\
5 9.004402557594164\\
10 9.004402557594164\\
15 9.004402557594164\\
20 9.004402557594164\\
25 9.004402557594164\\
30 9.004402557594164\\
};
\addlegendentry{$\Gamma_\text{EGT}(\mathbf{h_{10}})$};

\addplot [color=blue, line width=1.2pt, dashed, mark=o, mark size=1.5pt, mark options={solid, fill=blue}, mark repeat=1]
  table[row sep=crcr]{%
4 4.642509347717136\\
7 6.576790319051941\\
11 7.546307570187841\\
14 8.109565486093564\\
18 8.466985738895689\\
21 8.694097893723056\\
};
\addlegendentry{$\Gamma_\text{CP}(\mathbf{h_{10}})$};

\addplot [color=darkbrown, line width=1.2pt, loosely dotted, mark=o, mark size=1.5pt, mark options={solid, fill=darkbrown}, mark repeat=1]
  table[row sep=crcr]{%
9 4.742177268983696\\
18 7.470992873725313\\
24 8.29643804586263\\
};
\addlegendentry{$\Gamma_\text{PSK}(\mathbf{h_{10}})$};

\addplot [color=red, line width=1.2pt, dashdotted, mark=o, mark size=1.5pt, mark options={solid, fill=red}, mark repeat=1]
  table[row sep=crcr]{%
4 4.798044995946785\\
7 5.630647612061725\\
11 5.637988953221029\\
14 5.638049957815642\\
18 5.638050446442322\\
21 5.638050450236554\\
};
\addlegendentry{$\Gamma_\text{DFT}(\mathbf{h_{10}})$};



\end{axis}

\end{tikzpicture}%
    \end{minipage}%
  }
  \hspace{0.75cm}
  \hfill
  \subfloat[High correlation\label{fig:miso_exp}]{%
    \begin{minipage}[t]{0.45\textwidth}
      \begin{tikzpicture}
\definecolor{black}{rgb}{0, 0.0, 0}%
\definecolor{blue}{rgb}{0, 0, 1}%
\definecolor{red}{rgb}{1, 0, 0}%



\begin{axis}[
font=\footnotesize,
width=0.7\columnwidth,
height=4.5cm,
scale only axis,
xmin=1,
xmax=25,
xtick = {0,5,10,...,25},
xlabel={Feedback bits ($B$)},  
xmajorgrids,
ymin= 4,
ymax= 10,
ytick = {4, 5, ..., 10},
ylabel={$\Gamma_{\text{ave}}[\text{dB}]$},
ylabel near ticks,
ymajorgrids,
grid style={gray!80},
legend style={
  font=\small, 
  at={(0.5,1.02)}, 
  anchor=south, 
  draw=black, 
  fill=white, 
  legend cell align=center,
  /tikz/column 1/.append style={column sep=5pt},
  /tikz/column 2/.append style={column sep=5pt},
  /tikz/column 3/.append style={column sep=5pt},
},
legend columns=2,
]

\draw[->, thick, >=stealth] (axis cs:18,1.05) -- (axis cs:20,1.0);
\node[font=\scriptsize, anchor=east] at (axis cs:18,1.05) {EGT};



\addplot [color=black, line width=1.5pt, densely dotted, mark=triangle, mark size=1.5pt, mark options={solid, fill=blue}, mark repeat=1]
  table[row sep=crcr]{%
0 9.131543105573549\\
5 9.131543105573549\\
10 9.131543105573549\\
15 9.131543105573549\\
20 9.131543105573549\\
25 9.131543105573549\\
30 9.131543105573549\\
};
\addlegendentry{$\Gamma_\text{EGT}(\mathbf{h_{10}})$};

\addplot [color=blue, line width=1.2pt, dashed, mark=triangle, mark size=1.5pt, mark options={solid, fill=blue}, mark repeat=1]
  table[row sep=crcr]{%
4 6.774738739759768\\
7 7.312920988961611\\
11 7.851157953447258 \\
14 8.303309866852421\\
18 8.580675130944847\\
21 8.810153870248826\\
};
\addlegendentry{$\Gamma_\text{CP}(\mathbf{h_{10}})$};

\addplot [color=darkbrown, line width=1.2pt, loosely dotted, mark=triangle, mark size=1.5pt, mark options={solid, fill=darkbrown}, mark repeat=1]
  table[row sep=crcr]{%
9 5.278092795776010\\
18 7.630015997138571\\
24 8.412942642605682\\
};
\addlegendentry{$\Gamma_\text{PSK}(\mathbf{h_{10}})$};

\addplot [color=red, line width=1.2pt, dashdotted, mark=triangle, mark size=1.5pt, mark options={solid, fill=red}, mark repeat=1]
  table[row sep=crcr]{%
4 7.098711459268152\\
7 7.920365225848489\\
11 7.927718265628654\\
14 7.92777827578498\\
18 7.927778743176692\\
21 7.927778747297624\\
};
\addlegendentry{$\Gamma_\text{DFT}(\mathbf{h_{10}})$};



\end{axis}

\end{tikzpicture}%
    \end{minipage}%
  }
  \caption{Average beamforming gain $\Gamma_{\text{ave}}[\text{dB}]$ 
           versus feedback bits $B$ for correlated Rayleigh fading in 
           MISO system ($N_{\text{T}} = 10$)
           with (a) $\rho_{\text{TX}} = 0.2$ and (b) $\rho_{\text{TX}} = 0.75.$ The $\CCP$ codebook approaches the EGT baseline in both regimes without codebook rotation. DFT saturates below EGT at low correlation but improves at high correlation, consistent with its ULA-specific design. PSK remains largely insensitive to increasing feedback bits.}
  \label{fig:miso_rayleigh_exp}
\end{figure*}


\subsubsection{3GPP CDL-A}
\label{ssec:miso_cdl}
{The CDL-A channel is generated following the 3GPP TR 38.901 narrowband
CDL model~\cite{3GPP_TR_38901}, using the 23-cluster
NLOS power profile and angles of departure and arrival from
Table~7.7.1-1. The channel is given by
$$\mathbf{h} = \sum_{n=1}^{N_{\text{cl}}} \sqrt{P_n}\, g_n\,
\mathbf{a}_{\text{tx}}^{\text{T}}(\phi_n^{\text{tx}}),$$
where $g_n \sim \mathcal{CN}(0,1)$ are independent per-cluster fading
coefficients, $P_n$ are the normalized cluster powers, and
$\mathbf{a}_{\text{tx}}(\phi) = \frac{1}{\sqrt{N_{\text{T}}}}\bigl[1,\, e^{j\pi\sin\phi},\,
\ldots,\, e^{j\pi(N_{\text{T}}-1)\sin\phi}\bigr]^{\text{T}}$ is the ULA steering vector
with half-wavelength ULA antenna spacing.}
 \autoref{Res: cdl-a miso} presents the average beamforming gain versus the feedback bits $B$ for a 
MISO system with $N_{\text{T}} = 10$. The $\CCP$ codebook approaches $\Gamma_{\text{EGT}}$ smoothly with increasing $B$, while DFT plateaus below the EGT baseline, and PSK improves slowly, remaining below CP
across all feedback rates.
\begin{figure}[!htbp]
\centering
  {\begin{tikzpicture}
\definecolor{black}{rgb}{0, 0.0, 0}%
\definecolor{blue}{rgb}{0, 0, 1}%
\definecolor{red}{rgb}{1, 0, 0}%



\begin{axis}[
font=\footnotesize,
width=0.65\columnwidth,
height=4.5cm,
scale only axis,
xmin=1,
xmax=25,
xtick = {0,5,10,...,25},
xlabel={Feedback bits ($B$)},  
xmajorgrids,
ymin= 5,
ymax= 10,
ytick = {5, 6, ..., 10},
ylabel={$\Gamma_{\text{ave}}[\text{dB}]$},
ylabel near ticks,
ymajorgrids,
grid style={gray!80},
legend style={
  font=\small, 
  at={(0.5,1.02)}, 
  anchor=south, 
  draw=black, 
  fill=white, 
  legend cell align=center,
  /tikz/column 1/.append style={column sep=5pt},
  /tikz/column 2/.append style={column sep=5pt},
  /tikz/column 3/.append style={column sep=5pt},
},
legend columns=2,
]

\draw[->, thick, >=stealth] (axis cs:18,1.05) -- (axis cs:20,1.0);
\node[font=\scriptsize, anchor=east] at (axis cs:18,1.05) {EGT};

\addplot [color=black, line width=1.5pt, densely dotted, mark=triangle, mark size=1.5pt, mark options={solid, fill=blue}, mark repeat=1]
  table[row sep=crcr]{%
0 9.483572152191089\\
5 9.483572152191089\\
10 9.483572152191089\\
15 9.483572152191089\\
20 9.483572152191089\\
25 9.483572152191089\\
30 9.483572152191089\\
};
\addlegendentry{$\Gamma_\text{EGT}(\mathbf{h_{10}})$};

\addplot [color=blue, line width=1.2pt, dashed, mark=triangle, mark size=1.5pt, mark options={solid, fill=blue}, mark repeat=1]
  table[row sep=crcr]{%
4 6.085001211183005\\
7 6.892544327982407\\
11 7.772890588243549 \\
14 8.467385679447258\\
18 8.933140308605747\\
};
\addlegendentry{$\Gamma_\text{CP}(\mathbf{h_{10}})$};

\addplot [color=darkbrown, line width=1.2pt, dotted, mark=triangle, mark size=1.5pt, mark options={solid, fill=darkbrown}, mark repeat=1]
  table[row sep=crcr]{%
9 4.973923786039452\\
18 7.902739277133190\\
24 8.745099193300677\\
};
\addlegendentry{$\Gamma_\text{PSK}(\mathbf{h_{10}})$};

\addplot [color=red, line width=1.2pt, dashdotted, mark=triangle, mark size=1.5pt, mark options={solid, fill=red}, mark repeat=1]
  table[row sep=crcr]{%
4 6.969911055409883\\
7 8.449397965073413\\
11 8.458871606363177\\
14 8.458936311564319\\
18 8.458936861756499\\
};
\addlegendentry{$\Gamma_\text{DFT}(\mathbf{h_{10}})$};



\end{axis}

\end{tikzpicture}%
}
  \caption{Average beamforming gain $\Gamma_{\text{ave}}[\text{dB}]$ 
           versus feedback bits $B$ for CDL-A fading channel in 
           MISO system with $N_{\text{T}} = 10$. The $\CCP$ codebook approaches EGT baseline smoothly with increasing $B$, while DFT plateaus significantly below it and PSK improves slowly.}
  \label{Res: cdl-a miso}  
\end{figure}

\begin{remark}
    The performance of DFT codebooks is intrinsically tied to the Toeplitz structure of ULA correlation matrices. DFT codewords are well-aligned with dominant channel directions in highly correlated ULA settings but saturate below the EGT baseline in rich scattering environments. PSK codebooks, while satisfying the constant envelope constraint, achieve lower beamforming gain that is largely insensitive to increasing feedback bits. In contrast, $\CCP$ codebooks are constructed using Grassmannian line packing without assumptions on channel geometry, allowing robust performance across all channel models.
\end{remark}

\subsection{MIMO system}
\subsubsection{Iterative algorithm simulation for MIMO system with $N_{\text{R}}=2$}
\label{sec: Iterative algorithm simulations}

The following $\CCP$ codebook design simulations for MIMO system with $N_{\text{R}}=2$ are evaluated against the EGT gain baseline computed using \autoref{alg:Iterative EGT update}. The iterative algorithm is implemented with parameters as shown in \autoref{tab:Iterative algorithm parameters Rayleigh} for the Rayleigh fading channel. 
The parameter $b$ bits represents $2^b$ levels of phase quantization chosen for each of the $N_{\text{T}}$ phases of the channel. The parameter $N$ signifies the number of searches we perform for each co-ordinate of the beamforming vector to finally conclude the closest realization that provides the maximum EGT gain. The $N_{\textup{sim}}$ is the number of Monte Carlo realizations for each type of fading channel. 
{The chosen parameters $b = 8,N = 10, N_{\text{sim}}=300$ were verified to yield converged $\Gamma_{\text{EGT}}(\mathbf{H})$ values. Increasing either parameter did not meaningfully alter the baseline results for different fading profiles.} In the next subsections, the $\CCP$ gain from simulations is compared against the baseline established by~\autoref{alg:Iterative EGT update}.
\begin{center}
\begin{table}[t]
\centering
\caption{Design parameters chosen to implement the~\autoref{alg:Iterative EGT update} to establish the EGT baseline for MIMO system with $N_{\text{R}}=2$ for Rayleigh fading channel.}
\label{tab:Iterative algorithm parameters Rayleigh}
\begin{tabular}{|l|c|c|c|c|}
\hline
{Channel} & {$b$} & {$N$} & {$N_{\textup{sim}}$} & $\Gamma_{\textup{EGT}}[\text{dB}]$ \\
\hline
$\mathbf{H_{2 \times 4}}$ & 8 & 10 & 300 & 7.2320\\
\hline
$\mathbf{H_{2 \times 6}}$ & 8 & 10 & 300 & 8.6502 \\
\hline
$\mathbf{H_{2 \times 10}}$ & 8 & 10 & 300 & 10.4464 \\
\hline
\end{tabular}
\end{table}
\end{center}



\subsubsection{Rayleigh fading MIMO system}
\label{sec:Rayleigh fading_Nr_2}

\begin{figure}[!htbp]
\centering
  {\begin{tikzpicture}
\definecolor{black}{rgb}{0, 0.0, 0}%
\definecolor{blue}{rgb}{0, 0, 1}%
\definecolor{red}{rgb}{1, 0, 0}%

\definecolor{apcolor}{HTML}{7E2F8E}

\begin{axis}[
font=\footnotesize,
width=0.65\columnwidth,
height=4.5cm,
scale only axis,
xmin=1,
xmax=6,
xtick = {1,2,3,...,6},
xlabel={CP code dimension ($k$)},  
xmajorgrids,
ymin= 5,
ymax= 11,
ytick = {5, 6, ..., 12},
ylabel={$\Gamma_{\text{ave}}[\text{dB}]$},
ylabel near ticks,
grid style={gray!80},
ymajorgrids,
legend style={font=\small, at={(-0.125,1.03)},anchor=south west, draw=black,
fill=white,legend cell align=left},
legend columns = 3
]





\addplot [color=red, densely dotted, mark=star, mark options = {solid}, line width=1.2pt]
  table[row sep=crcr]{%
1 7.232063659621407\\
2 7.232063659621407\\
3 7.232063659621407\\
4 7.232063659621407\\
};
\addlegendentry{$\Gamma_\text{EGT}(\mathbf{H_{2 \times 4}})$}

\addplot [color=apcolor, dashed, mark=o, mark options = {solid}, line width=1.5pt]
  table[row sep=crcr]{%
1 5.579720839042953\\
2 6.573950092549018\\
3 7.087778641997556\\
};
\addlegendentry{$\Gamma_\text{AP}(\mathbf{H_{2 \times 4}})$};

\addplot [color=black, densely dotted, mark=diamond, mark options = {solid}, line width=1.2pt]
  table[row sep=crcr]{%
1 8.650256003416851\\
2 8.650256003416851\\
3 8.650256003416851\\
4 8.650256003416851\\
5 8.650256003416851\\
6 8.650256003416851\\
};
\addlegendentry{$\Gamma_\text{EGT}(\mathbf{H_{2 \times 6}})$};

\addplot [color=blue, densely dotted, mark=triangle, mark options = {solid}, line width=1.2pt]
  table[row sep=crcr]{%
1 10.446439207410315\\
2 10.446439207410315\\
3 10.446439207410315\\
4 10.446439207410315\\
5 10.446439207410315\\
6 10.446439207410315\\
7 10.446439207410315\\
};
\addlegendentry{$\Gamma_\text{EGT}(\mathbf{H_{2 \times 10}})$};

\addplot [color=red, dash dot, mark=asterisk, mark options = {solid}, line width=1.2pt]
  table[row sep=crcr]{%
1 5.600309499\\
2 6.544468350\\
3 6.948248554 \\
4 6.948248554\\
};
\addlegendentry{$\Gamma_\text{CP}(\mathbf{H_{2 \times 4}})$};

\addplot [color=black, dash dot, mark=diamond, mark options = {solid}, line width=1.2pt]
  table[row sep=crcr]{%
1 6.082872612\\
2 7.317166317\\
3 7.937219220 \\
4 8.279778688\\
5 8.438019714\\
6 8.438019715\\
};
\addlegendentry{$\Gamma_\text{CP}(\mathbf{H_{2 \times 6}})$};

\addplot [color=blue, solid, dash dot, mark=triangle, mark options = {solid}, line width=1.2pt]
  table[row sep=crcr]{%
1 6.542752451\\
2 8.289080548\\
3 9.114737039\\
4 9.683829348\\
5 9.993889866\\
6 10.214797739549912\\
};
\addlegendentry{$\Gamma_\text{CP}(\mathbf{H_{2 \times 10}})$};

\node[anchor=west] at (3.05,6.6){\textcolor{red}{$N_T = 4$}};
\node[anchor=north] at (4.5,8.0){\textcolor{black}{$N_T = 6$}};
\node[anchor=north] at (5.5,9.8){\textcolor{blue}{$N_T = 10$}};

\end{axis}

\end{tikzpicture}%
}
  \caption{Average beamforming gain $\Gamma_{\text{ave}}[\text{dB}]$ as a 
  function
  of the $\CCP$ code dimension $k$ for Rayleigh fading channel in 
  MIMO system. The $\CCP$ gain increases smoothly with $k$ and approaches the
EGT baseline (computed via~\autoref{alg:Iterative EGT update}) for all three antenna configurations, confirming that the
MISO convergence behavior extends to MIMO.}
  \label{Res: Rayleigh fading Nr_2}  
\end{figure}

\autoref{Res: Rayleigh fading Nr_2} shows the average beamforming gain versus the $\CCP$ code dimension $k$ for a $2 \times N_{\text{T}}$ MIMO system under Rayleigh fading, where the EGT baseline is computed via Algorithm~1 with the parameters listed in~\autoref{tab:Iterative algorithm parameters Rayleigh}. As in the MISO case, the $\CCP$ beamforming gain $\Gamma_{\text{CP}}(\mathbf{H})$ increases smoothly with $k$ and approaches $\Gamma_{\text{EGT}}(\mathbf{H})$ as $k$ approaches $n$, confirming that the trade-off between feedback rate and beamforming performance extends to the MIMO setting. The behavior is consistent across $N_{\text{T}} \in \{4, 6, 10\}$, with the gap to the EGT baseline narrowing at higher code rates in each case. Performance is evaluated over 300 Monte Carlo realizations of the channel $\mathbf{H}$. 
{
$\Gamma_{\text{AP}}(\mathbf{H}_{2\times4})$ is included 
for $N_{\text{T}} = 4$, $k \leq 3$, and remains within $0.14$~dB of $\Gamma_{\text{CP}}(\mathbf{H}_{2\times4})$, confirming that the packing advantage  of AP does not translate to a meaningful gain over the structured $\CCP$ codebook.}

\subsubsection{3GPP CDL-A Channel}
{The CDL-A channel is generated following the 3GPP TR 38.901 
narrowband model~\cite{3GPP_TR_38901} with $23$ clusters and 
half-wavelength ULA antenna spacing.
For the MIMO case with $N_{\text{R}} = 2$, 
the channel matrix is given by
$$\mathbf{H} = \sum_{n=1}^{N_{\text{cl}}} \sqrt{P_n}\, g_n\,
\mathbf{a}_{\text{rx}}(\phi_n^{\text{rx}})\,
\mathbf{a}_{\text{tx}}^{\text{T}}(\phi_n^{\text{tx}}),$$
where $\mathbf{a}_{\text{tx}}(\phi)$ is the ULA steering 
vector, and 
$\mathbf{a}_{\text{rx}}(\phi) = \frac{1}{\sqrt{2}}
\bigl[1,\, e^{j\pi\sin\phi}\bigr]^{\text{T}}$ is the receive 
steering vector with $N_{\text{R}} = 2$ antennas.}
\autoref{Res: cdl_mimo} presents the average beamforming gain versus feedback bits $B$ for the CDL-A channel in a $2\times10$ MIMO system, where the EGT baseline $\Gamma_{\text{EGT}}(\mathbf{H})$ is computed via~\autoref{alg:Iterative EGT update} to be $11.9425\text{ dB}$. The $\CCP$ codebook approaches the EGT baseline smoothly with increasing $B$, while PSK improves at a significantly slower rate, as constellation-based codebooks are not designed with Grassmannian line packing, limiting their 
beamforming resolution regardless of the feedback rate. The DFT codebook achieves higher gain than $\CCP$ at low feedback rates, since its beams are ULA steering vectors that align
closely with the dominant angular clusters of the CDL-A channel
even under coarse quantization. As $B$ increases, however, DFT
plateaus below the EGT baseline due to its fixed, geometry-specific
beam grid. The $\CCP$ codebook, in contrast, continues to
improve with increasing $B$ and surpasses DFT,
approaching the EGT baseline.
\begin{figure}[!htbp]
\centering
  {\begin{tikzpicture}
\definecolor{black}{rgb}{0, 0.0, 0}%
\definecolor{blue}{rgb}{0, 0, 1}%
\definecolor{red}{rgb}{1, 0, 0}%



\begin{axis}[
font=\footnotesize,
width=0.65\columnwidth,
height=4.5cm,
scale only axis,
xmin=1,
xmax=25,
xtick = {0,5,10,...,25},
xlabel={Feedback bits ($B$)},  
xmajorgrids,
ymin= 7,
ymax= 13,
ytick = {7,8,  ..., 13},
ylabel={$\Gamma_{\text{ave}}[\text{dB}]$},
ylabel near ticks,
ymajorgrids,
grid style={gray!80},
legend style={
  font=\small, 
  at={(0.5,1.02)}, 
  anchor=south, 
  draw=black, 
  fill=white, 
  legend cell align=center,
  /tikz/column 1/.append style={column sep=5pt},
  /tikz/column 2/.append style={column sep=5pt},
  /tikz/column 3/.append style={column sep=5pt},
},
legend columns=2,
]



\addplot [color=black, line width=1.5pt, dashed, mark=triangle, mark size=1.5pt, mark options={solid, fill=blue}, mark repeat=1]
  table[row sep=crcr]{%
0 11.942509853353922\\
5 11.942509853353922\\
10 11.942509853353922\\
15 11.942509853353922\\
20 11.942509853353922\\
25 11.942509853353922\\
30 11.942509853353922\\
};
\addlegendentry{$\Gamma_\text{EGT}(\mathbf{H_{2 \times 10}})$};

\addplot [color=blue, line width=1.2pt, dashed, mark=triangle, mark size=1.5pt, mark options={solid, fill=blue}, mark repeat=1]
  table[row sep=crcr]{%
4 9.094790739357666\\
7 9.528555525720492\\
11 10.260778882966136 \\
14 10.987035194252696\\
18 11.49317634602085\\
21 11.686020537208934\\
};
\addlegendentry{$\Gamma_\text{CP}(\mathbf{H_{2 \times 10}})$};

\addplot [color=darkbrown, line width=1.2pt, dotted, mark=triangle, mark size=1.5pt, mark options={solid, fill=darkbrown}, mark repeat=1]
  table[row sep=crcr]{%
9 7.786464172033684\\
18 10.485169716183105\\
24 11.309588798054008\\
};
\addlegendentry{$\Gamma_\text{PSK}(\mathbf{H_{2 \times 10}})$};

\addplot [color=red, line width=1.2pt, dashdotted, mark=triangle, mark size=1.5pt, mark options={solid, fill=red}, mark repeat=1]
  table[row sep=crcr]{%
4 9.891407521481138\\
7 11.373475330895985\\
11 11.382054932306279\\
14 11.382125960869269\\
18 11.382127835468124\\
21 11.382127632316173\\
};
\addlegendentry{$\Gamma_\text{DFT}(\mathbf{H_{2 \times 10}})$};



\end{axis}

\end{tikzpicture}%
}
  \caption{Average beamforming gain $\Gamma_{\text{ave}}[\text{dB}]$ 
           versus feedback bits $B$ for CDL-A channel in 
           MIMO system. The $\CCP$ codebook consistently approaches EGT baseline while the DFT plateaus below the EGT baseline and PSK improves slowly. }
  \label{Res: cdl_mimo}  
\end{figure}

\begin{remark}
    In line-of-sight (LoS) dominated channels such as CDL-D~\cite{3GPP_TR_38901} and Rician fading~\cite{heath2018foundations} with high $\kappa$-factor, additional simulations 
    {(omitted for brevity), consistent with LoS behavior,} indicate that the channel energy concentrates in a single dominant direction. In such regimes, both CP and DFT approach the EGT baseline, as any well-designed codebook can identify the dominant direction with sufficient feedback bits. This confirms that CP performs robustly across the full spectrum from rich scattering, where its Grassmannian packing advantage is most pronounced, to LoS channels, where it matches geometry-specific designs without any channel model knowledge.
\end{remark}
The simulation results across Rayleigh, correlated Rayleigh, and CDL channel models for both MISO and MIMO systems demonstrate that the $\CCP$ codebook consistently approaches the EGT baseline with increasing feedback bits. Unlike DFT codebooks, whose performance depends on channel correlation and array geometry, and PSK codebooks, which remain largely insensitive to increasing feedback rates, the $\CCP$ codebook provides robust beamforming performance across all channel conditions without requiring codebook rotation or channel-model-specific adaptation.
\section{Conclusion}
\label{sec:Future Work}

In this paper, we studied the problem of precoding codebook design for limited-feedback MIMO systems under the equal-gain transmission constraint. We showed that $\CCP$ codes can be used to construct structured precoding codebooks with constant-envelope codewords and low storage complexity. For MISO systems, we derived bounds on the mean squared quantization error and showed that the beamforming gain approaches the EGT baseline as the code rate increases. For MIMO systems with multiple receive antennas, we developed an iterative algorithm to compute the EGT baseline and demonstrated through simulations that $\CCP$ codebooks approach this baseline across a range of channel models, including i.i.d. Rayleigh, spatially correlated, and standardized CDL channels. We also compared $\CCP$ codebooks with practical benchmarks, including DFT and PSK codebooks, and showed that $\CCP$ provides robust performance across different channel conditions without requiring channel statistics, codebook rotation, or array-specific design. 
{Comparison with the AP Grassmannian codebook, a numerically optimized benchmark, confirms that the structured packing of $\CCP$ codes achieves near-optimal Grassmannian quantization, with a gap of less than 0.15~dB for a four-transmit-antenna configuration.}

These results suggest that algebraically structured codebooks designed via Grassmannian packing provide a viable alternative to geometry-dependent codebooks currently used in practice. As antenna arrays become larger and channel environments become more heterogeneous, codebook designs that do not rely on specific channel statistics or array geometries may offer greater robustness and flexibility. The results in this paper indicate that structured Grassmannian packing, combined with constant-envelope constraints, can provide a useful framework for codebook design in such scenarios.

{Additionally, all results in this paper assume perfect CSIR. The relative performance of $\CCP$, DFT, and PSK codebooks under imperfect CSIR remains an open question.} Several directions remain for future work. These include extending the proposed framework to higher-rank precoding and multi-stream transmission, adapting the approach to near-field and extremely large array regimes, and developing lower-complexity search and decoding algorithms that exploit the algebraic structure of $\CCP$ codes. More broadly, the connection between algebraic coding theory and Grassmannian quantization explored in this work suggests that additional algebraic constructions may lead to new families of structured precoding codebooks for next-generation wireless systems.





\end{document}